\documentclass[12pt]{article}

\def\includefigs{\let\ifincfigs=\iftrue}
\def\noincludefigs{\let\ifincfigs=\iffalse}
\includefigs

\newbox\epsfvertlab
\newbox\epsfhorlab
\newbox\epsffiglab

\newdimen\epsfvlabsize
\newdimen\scott

\def\setvlabel#1{\setbox\epsfvertlab=\vbox{\hbox{#1}}}%
\def\sethlabel#1{\setbox\epsfhorlab=\vbox{\hbox{#1}}}%
\def\figlab#1 #2 #3{\setbox\epsffiglab=\vbox to 0pt{%
\ifvoid\epsffiglab\else\box\epsffiglab\fi\vss\hbox to 0pt{\raise #2 \hbox{\hskip #1 #3}\hss}}}

\newdimen\fighor
\newdimen\figver
\newbox\rotbox
\long\def\lrlap#1{\hbox to 0pt{#1\hss}}
\long\def\verttex#1#2#3{{\fighor = #1\figver = #2\vbox to \figver{\vss%
\hbox to \fighor{\hfill\hsize=\fighor%
\lrlap{\rotstart{-90 rotate}\vbox to \fighor{#3\vfil}\rotfinish}}}}}

\def\dvipsvspec#1{\special{ps:#1}}
\def\dvipsrotstart#1{\dvipsvspec{gsave currentpoint currentpoint translate
   #1 neg exch neg exch translate}}
\def\dvipsrotfinish{\dvipsvspec{currentpoint grestore moveto}}

\def\rotstart#1{\dvipsrotstart{#1}}
\def\rotfinish{\dvipsrotfinish}

\def\epsfsetlab{%
\ifvoid\epsfvertlab%
\else%
\verttex{\epsfvlabsize}{\epsfysize}%
{\hbox to \epsfysize{\hss\box\epsfvertlab\hss}}%
\fi%
\ifvoid\epsfhorlab%
\else%
\scott=\epsfxsize%
\advance\scott by \epsfvlabsize%
\rlap{\vtop{\hrule height0pt\hbox to \scott{\hss\box\epsfhorlab\hss}}}%
\fi%
}

\def\epsfsetover{\ifvoid\epsffiglab\else\box\epsffiglab\fi}

\newread\epsffilein    
\newif\ifepsffileok    
\newif\ifepsfbbfound   
\newif\ifepsfverbose   
\newdimen\epsfxsize    
\newdimen\epsfysize    
\newdimen\epsftsize    
\newdimen\epsfrsize    
\newdimen\epsftmp      
\newdimen\pspoints     
\def\epsfbox#1{
   \ifvoid\epsfvertlab%
   \else\epsfvlabsize=\ht\epsfvertlab \advance\epsfvlabsize by \dp\epsfvertlab\fi%
   \leavevmode\global\def\epsfllx{72}\global\def\epsflly{72}%
   \global\def\epsfurx{540}\global\def\epsfury{720}%
   \def\lbracket{[}\def\testit{#1}\ifx\testit\lbracket
   \let\next=\epsfgetlitbb\else\let\next=\epsfnormal\fi\next{#1}}%
\def\epsfgetlitbb#1#2 #3 #4 #5]#6{\epsfgrab #2 #3 #4 #5 .\\%
   \epsfsetgraph{#6}}%
\def\epsfnormal#1{\epsfgetbb{#1}\epsfsetgraph{#1}}%
\def\epsfgetbb#1{%
\openin\epsffilein=#1
\ifeof\epsffilein\errmessage{I couldn't open #1, will ignore it}\else
   {\epsffileoktrue \chardef\other=12
    \def\do##1{\catcode`##1=\other}\dospecials \catcode`\ =10
    \loop
       \read\epsffilein to \epsffileline
       \ifeof\epsffilein\epsffileokfalse\else
          \expandafter\epsfaux\epsffileline:. \\%
       \fi
   \ifepsffileok\repeat
   \ifepsfbbfound\else
    \ifepsfverbose\message{No bounding box comment in #1; using defaults}\fi\fi
   }\closein\epsffilein\fi}%
\def\epsfsetgraph#1{%
   \epsfrsize=\epsfury\pspoints
   \advance\epsfrsize by-\epsflly\pspoints
   \epsftsize=\epsfurx\pspoints
   \advance\epsftsize by-\epsfllx\pspoints
   \epsfxsize\epsfsize\epsftsize\epsfrsize
   \ifnum\epsfxsize=0 \ifnum\epsfysize=0
      \epsfxsize=\epsftsize \epsfysize=\epsfrsize
     \else\epsftmp=\epsftsize \divide\epsftmp\epsfrsize
       \epsfxsize=\epsfysize \multiply\epsfxsize\epsftmp
       \multiply\epsftmp\epsfrsize \advance\epsftsize-\epsftmp
       \epsftmp=\epsfysize
       \loop \advance\epsftsize\epsftsize \divide\epsftmp 2
       \ifnum\epsftmp>0
          \ifnum\epsftsize<\epsfrsize\else
             \advance\epsftsize-\epsfrsize \advance\epsfxsize\epsftmp \fi
       \repeat
     \fi
   \else\epsftmp=\epsfrsize \divide\epsftmp\epsftsize
     \epsfysize=\epsfxsize \multiply\epsfysize\epsftmp   
     \multiply\epsftmp\epsftsize \advance\epsfrsize-\epsftmp
     \epsftmp=\epsfxsize
     \loop \advance\epsfrsize\epsfrsize \divide\epsftmp 2
     \ifnum\epsftmp>0
        \ifnum\epsfrsize<\epsftsize\else
           \advance\epsfrsize-\epsftsize \advance\epsfysize\epsftmp \fi
     \repeat     
   \fi
   \ifepsfverbose\message{#1: width=\the\epsfxsize, height=\the\epsfysize}\fi
   \epsftmp=10\epsfxsize \divide\epsftmp\pspoints
   \epsfsetlab%
   \ifincfigs%
     \vbox to\epsfysize{\vfil\hbox to\epsfxsize{%
        \includegraphics{#1}%
        \epsfsetover\hfil}}%
   \else%
     \epsfsetover%
     \vbox to\epsfysize{\hrule\vss\hbox to\epsfxsize{\vrule height
                        \epsfysize\hfil\vrule}\vss\hrule}%
   \fi%
\epsfxsize=0pt\epsfysize=0pt}%

{\catcode`\%=12 \global\let\epsfpercent=
\long\def\epsfaux#1#2:#3\\{\ifx#1\epsfpercent
   \def\testit{#2}\ifx\testit\epsfbblit
      \epsfgrab #3 . . . \\%
      \epsffileokfalse
      \global\epsfbbfoundtrue
   \fi\else\ifx#1\par\else\epsffileokfalse\fi\fi}%
\def\epsfgrab #1 #2 #3 #4 #5\\{%
   \global\def\epsfllx{#1}\ifx\epsfllx\empty
      \epsfgrab #2 #3 #4 #5 .\\\else
   \global\def\epsflly{#2}%
   \global\def\epsfurx{#3}\global\def\epsfury{#4}\fi}%
\def\epsfsize#1#2{\epsfxsize}
\def\ifspace{\ifcat\issp.\else~\fi}
\def\tspace{\futurelet\issp\ifspace}
\def\a{({\it a\kern 1pt})\tspace}
\def\b{({\it b\kern 1pt})\tspace}
\def\c{({\it c\kern 1pt})\tspace}
\def\d{({\it d\kern 1pt})\tspace}
\def\e{({\it e\kern 1pt})\tspace}
\def\f{({\it f\kern 1pt})\tspace}
\def\g{({\it g\kern 1pt})\tspace}
\def\h{({\it h\kern 1pt})\tspace}
\def\i{({\it i\kern 1pt})\tspace}
\def\j{({\it j\kern 1pt})\tspace}
\def\abc#1{({\it #1\kern 1pt})\tspace}

\newcount\ndots
\def\drawline#1#2{\raise 2.5pt\vbox{\hrule width #1pt height #2pt}}

\def\trian{\raise 1.25pt\hbox{$\scriptscriptstyle\triangle$}\nobreak\ }
\def\solidtrian{\raise 1.25pt
\hbox to 3bp{
\hfill}\nobreak\ }
\def\dsolidtrian{\raise 1.25pt
\hbox to 3bp{
\hfill}\nobreak\ }
\def\soliddiamond{\raise 1.25pt
\hbox to 4bp{
\hfill}\nobreak\ }

\def\square{${\vcenter{\hrule height .4pt 
              \hbox{\vrule width .4pt height 3pt \kern 3pt \vrule width .4pt}
          \hrule height .4pt}}$\nobreak\ }

\def\plus{\raise 1.25pt \hbox{$\scriptscriptstyle +$}\nobreak\ }
\def\x{\raise 1.25pt \hbox{$\scriptscriptstyle \times$}\nobreak\ }
\def\legendtable#1{\vbox{\baselineskip=10pt\tabskip=0pt\let\\=\cr\halign{\hfil##\hskip 3pt&##\hfil\cr#1\crcr}}}
\def\lllegend#1 #2 #3{\figlab {#1} {#2} {\legendtable{#3}}}
\def\lrlegend#1 #2 #3{\figlab {#1} {#2} {\llap{\legendtable{#3}}}}
\def\ullegend#1 #2 #3{\figlab {#1} {#2} {\vtop{\hrule height 0pt\legendtable{#3}}}}
\def\urlegend#1 #2 #3{\figlab {#1} {#2} {\llap{\vtop{\hrule height 0pt\legendtable{#3}}}}}

\newdimen\xorigon
\newdimen\yorigon
\newdimen\scaleval
\newdimen\scaleorigon

\def\setxscale#1 #2 #3 #4 #5 {%
    \xorigon=#1\yorigon=#3%
    \scaleval=#2\advance\scaleval by -\xorigon%
    \tempdimen=#5 pt\advance\tempdimen by -#4pt%
    \divide\tempdimen by 1000%
    \divide\scaleval by \tempdimen%
    \scaleorigon=-#4pt\divide\scaleorigon by 1000%
    \multiply\scaleorigon by \scaleval}
\def\xtickup#1 #2{\tempdimen=#1pt\divide\tempdimen by 1000%
    \multiply\tempdimen by \scaleval\advance\tempdimen by \scaleorigon%
    \advance\tempdimen by \xorigon%
    \figlab {\tempdimen} {\yorigon} {\vbox {\hbox to 0pt{\hss #2\hss}%
        \baselineskip=8pt\lineskiplimit=-5pt%
        \hbox to 0pt{\hss \vrule height 3pt\hss}}}}
\def\xtickdown#1 #2{\tempdimen=#1pt\divide\tempdimen by 1000%
    \multiply\tempdimen by \scaleval\advance\tempdimen by \scaleorigon%
    \advance\tempdimen by \xorigon%
    \figlab {\tempdimen} {\yorigon} {\vbox to 0pt {\hbox to 0pt{\hss \vrule height 3pt\hss}%
        \nointerlineskip\vskip 3pt%
        \hbox to 0pt{\hss #2\hss}\vss}}}
\def\nofig#1#2{\leavevmode{\vbox {\hrule \hbox to #1{\vrule height #2 \hfill \vrule} \hrule}} }

\usepackage{graphicx}
\setlength{\topmargin}{-0.5 in}
\setlength{\textwidth}{6.5 in}
\setlength{\textheight}{9.0in}
\setlength{\oddsidemargin}{0.0in}
\begin{document}
\begin{center}
\begin{Large}
{\bf Utilitarian opacity model for aggregate particles in protoplanetary nebulae and exoplanet atmospheres}\\
\end{Large}
Jeffrey N. Cuzzi$^{1,*}$, 
Paul R. Estrada$^2$, and Sanford S. Davis$^1$\\
$^1$Space Science Division, Ames Research Center, NASA; $^2$SETI Institute
\vspace{0.5 in}
$^*$jeffrey.cuzzi@nasa.gov

{\bf Abstract}
\end{center}
As small solid grains grow into larger ones in protoplanetary nebulae, or in the cloudy atmospheres of exoplanets, they generally form porous aggregates rather than solid spheres. A number of previous studies have used highly sophisticated schemes to calculate opacity models for irregular, porous particles with size much smaller than a wavelength. However, mere growth itself can affect the opacity of the medium in far more significant ways than the detailed compositional and/or structural differences between grain constituents once aggregate particle sizes exceed the relevant wavelengths. This physics is not new; our goal here is to provide a model that provides physical insight and is simple to use in the increasing number of protoplanetary nebula evolution, and exoplanet atmosphere, models appearing in recent years, yet quantitatively captures the main radiative properties of mixtures of particles of arbitrary size, porosity, and composition. The model is a simple combination of effective medium theory with small-particle closed-form expressions, combined with suitably chosen transitions to geometric optics behavior. Calculations of wavelength-dependent emission and Rosseland mean opacity are shown and compared with Mie theory. The model's fidelity is very good in all comparisons we have made, except in cases involving pure metal particles or monochromatic opacities for solid particles with size comparable to the wavelength.

\vspace{0.25 in}
Astrophysical Journal Supplement; submitted June 14 2013, accepted December 3, 2013
\section{Introduction}

The total extinction opacity of a medium $\kappa_e$ (cm$^{2}$g$^{-1}$) is the ratio of its
volume extinction coefficient (cm$^{-1}$ of path) to its volume mass density. Equivalently, it is the cross-section per unit mass along a path. In protoplanetary nebulae, the relatively small fraction (by mass) of solid particles provides the bulk of the opacity (D'Alessio et al 1999, 2001) unless the temperature is so large (above 1500K) that common geological solids evaporate and molecular species dominate (Nakamoto and Nakagawa 1994, Ferguson et al 2005). In exoplanet atmospheres, which are generally much denser, the particle contributions are more situation-dependent (Marley et al 1999, Sudarsky et al 2000, 2003; Ackerman and Marley 2001, Tsuji 2002, Currie et al 2011, de Kok et al 2011, Madhusudhan et al 2011, Morley et al 2012, 2013; Vasquez et al 2013).

While typical interstellar grain size distributions (Mathis et al 1977, Draine and Lee 1984) have been assumed in many nebula and planetary atmospheric opacity models (Pollack et al 1985, 1994; D'Alessio et al 1999, 2001; Bodenheimer et al 2000, Dullemond et al 2002), coagulation models in the nebula context, going back to Weidenschilling (1988, 1997), and most recently Ormel and Okuzumi (2013), suggest that such small grains grow to 100$\mu$m size or larger on short timescales. This is because small and/or fluffy grains collide at very low relative velocity and stick readily. Other recent studies of grain growth in the nebula context include Brauer et al (2008), Zsom et al (2010), Schraepler et al (2012), and Birnstiel et al (2010, 2012) to give only a few examples.  In cloud {\it formation} models for gas giant planets (Movshovitz et al 2010) and for brown dwarfs and exoplanets (Helling et al 2001, Kaltenegger et al 2007, Kitzmann et al 2010, Zsom et al 2012). A review by Helling et al (2008) shows how exoplanet and brown dwarf clouds appear at different altitudes and with different particle sizes, formed from a variety of constituents from volatile ices to silicate or even iron metal. In a recent review, Marley et al (2013) illustrate how some materials condense as liquids (forming droplets that are  appropriate for the usual Mie theory models), while other materials condense as solids, forming irregular, porous aggregates as they grow. To our knowledge, no current exoplanet radiative transfer models have explored the implications of porous aggregate cloud particles. In this paper we present a simple way of accounting for radiative transfer involving aggregates of arbitrary size and porosity. Some giant planet atmosphere models (Podolak 2003, Movshovitz and Podolak 2008, Movshovitz et al 2010, and Helled and Bodenheimer 2011) and one other application (Amit and Podolak 2009) have already incorporated the opacity model we describe here. Using these opacity models, Movshovitz et al (2010) showed that realistic grain coagulation leads to smaller opacities, and shorter formation times, for gas giant planets than previously appreciated in the context of the Core-instability scenario (see section 5.4). 

Miyake and Nakagawa (1993), Pollack et al (1994), and D'Alessio et al (2001) give a few examples of the dramatic effect of particle growth on opacity across a broad range of sizes of interest in the nebula case. Because the aggregating constituents are not always liquid, the growing particles  can have a fairly open, highly porous, nature (Meakin and Donn 1988, Donn 1990, Dominik and Tielens 1997, Beckwith et al 2000, Dominik et al 2007, Blum 2010).  Growth beyond mm-cm sizes in protoplanetary nebulae remains an active area of study (see reviews by Cuzzi and Weidenschilling 2006 and Dominik et al 2007, and models by Brauer et al 2008, Zsom et al 2010, Hughes and Armitage 2012, Birnstiel et al 2010, 2012; and others), yet none of these latter models actually incorporates equally realistic and self-consistent treatments of opacity either in their radiative transfer or in predictions of Spectral Energy Distributions, and only Pollack et al (1985, 1994) and Miyake and Nakagawa (1993) have addressed the issue of large particle porosity in the implications for thermal radiative transfer. In both nebula and atmospheric cases, the vertical temperature structure is determined by the heating source (external and/or internal) and the opacity - usually expressed as a temperature-weighted mean such as the Rosseland mean opacity $\kappa_R$ (Pollack et al 1985, 1994; Henning and Stognienko 1996, Semenov et al 2003); the Planck mean can also be used for optically thin applications and is even simpler to calculate (Nakamoto and Nakagawa 1994). 

Interpretation of Spectral Energy Distributions, or intensity as a function of wavelength, requires an understanding of the wavelength-dependent opacity as well as the detailed vertical temperature distribution of the nebula or planet's atmosphere. Traditionally, regions sampled are thought to be optically thin, and particles poor scatterers being much smaller than the wavelength, so only the absorption part of the extinction opacity is needed (Miyake and Nakagawa 1993). Nebula gas masses have often been inferred by combining the mass of {\it particulates} (inferred from the observed emission and theoretical opacity) with the canonical mass ratio of roughly one percent for solids to total mass ({\it eg.} Andrews and Williams 2007, Williams and Cieza 2011). Theoretical analyses of observed mm-cm wavelength opacities at least since D'Alessio et al (2001) ({\it eg.} review by Natta et al 2007; also Isella et al 2009, Birnstiel et al 2010, Ricci et al 2010, and others) use full Mie theoretical analyses of particle opacities, which we show in section \ref{abs} is, in fact, essential when the particle size and wavelength are comparable. de Kok et al (2011) modeled emission spectra of exoplanets, where the primary spectral variation is that of the gas, but where the more gradually varying extinction properties of cloud particles still provide important variations across the near-IR passband because the thermal wavelengths are not extremely large compared to the particle size. 

{\it Overview:} Below we present a very simple, but general and flexible, model which captures the essential physics of opacity in ensembles of aggregate particles of arbitrary size and porosity, and can be easily made part of nebula or planetary atmosphere thermal structure and/or evolution models. We give some examples of how growth and porosity affect the opacity of realistic aggregate particles. The opacity model is simple to combine with any grain coagulation/aggregation model, and can be used to study the temporal/spatial dependence of the SED and Rosseland mean opacity $\kappa_R$ on varying particle size, density, porosity, and composition, or to calculate thermal equilibrium profiles. The model is not ideally suited to interpreting mm-cm wavelength spectral slopes in protoplanetary disks, where the dominant size is often found to be comparable to the wavelength. 

\section{Prior Work and overview of paper}

Draine and Lee (1984) studied small, independent particles ($r \ll \lambda$, where $r$ is particle radius and $\lambda$ is wavelength) in the dipole approximation;
Pollack et al (1985) modeled various size distributions of mineralogically
realistic, spherical, solid grains using exhaustively compiled refractive
indices and Mie scattering; Wright (1987) modeled grains of various fractal
dimension (basically, porosity) using the Discrete Dipole Approximation or DDA (Purcell and Pennypacker 1973, Draine and Goodman 1993). Wright (1987) 
showed that nonsphericity could cause a significant enhancement of opacity,
{\it if} grains had fairly large refractive indices. More recent work has elaborated upon this (Fabian et al 2001, Min et al 2008, Lindsay et al 2013), showing how this caveat applies to detailed spectral analysis of highly nonspherical particles  (elongated, flattened, or even having sharp edges) through strong absorption bands. Bohren and Huffman (1983, p. 140) also give constraints on how particle size and refractive index must both be considered, in cases where the refractive index is large. Our model is most suitable for roughly equidimensional particles with moderate refractive indices (high porosity and realistic compositional mixtures usually preclude high average indices). 

Ossenkopf (1991) studied the effect of metallic inclusions on equidimensional porous particles, emphasizing small particles and the role of randomly connected iron grains in leading to effective dipole-like structures. For the low volume densities of iron grains in protoplanetary nebula aggregates, at least for the abundances assumed here, the effect is small. Pollack et al (1985, 1994)
assumed materials which are cosmically abundant (with an eye towards
molecular cloud and protoplanetary nebula applications);  of these, water ice, silicates, and 
refractory organics have refractive indices that are too small for nonsphericity effects to be
especially important in the regime where particle size is smaller than a
wavelength. Other common materials, specifically iron metal and iron sulfide (Troilite, FeS) do have refractive indices that are high enough to show an effect, but nonsphericity was not treated by Pollack et al (1985, 1994).  

Regarding distributions of larger particles, Pollack et al (1985, 1994) and Miyake and Nakagawa (1993) calculated both monochromatic opacities and Rosseland Mean opacities, and D'Alessio et al (2001) calculated monochromatic opacities and employed them in a wide range of nebula structure models. Pollack et al (1985) showed selected results in the separate limits of large and small particles, but did not connect them. Mizuno et al (1988) used a combination of Rayleigh scattering for small particles and diffraction-augmented geometric optics ($Q_e = 2$) for large ones. Miyake and Nakagawa (1993) used a combination of full Mie calculations and geometrical optics, along with an {\it ad hoc} powerlaw particle size distribution, to demonstrate the effects of particle growth. To match confidently with the geometric optics limit, they carried out Mie calculations to impressively large values of $r/\lambda \sim 10^5$, with implications for array size and computational time that would be hard for the typical user to achieve even today, in multidimensional and/or evolutionary nebula or exoplanet atmosphere models. D'Alessio et al (2001) also seem to have used full Mie calculations for all their particles, up to 10cm radius, treating each material as a separate species. Pollack et al. (1985), Mizuno et al (1988), Miyake and Nakagawa (1993), and D'Alessio et al (2001) all assumed solid objects; Pollack et al (1994) looked briefly at particles of higher porosity, using an Effective Medium Theory (EMT; see section \ref{mixtures}) and Mie calculations. Rannou et al (1999) used a semi-empirical model of their own design to capture the properties of aggregates in certain size and compositional regimes. 

In this paper, we build on and generalize this prior work by showing how arbitrary {\it size}, {\it porosity}, and {\it compositional} distributions can be handled easily by users wanting to explore their own choices for grain properties, and simply enough to incorporate within evolutionary models or in iterative analysis of observed  Spectral Energy Distributions for applications where grain growth into sizeable aggregates has occurred. We show how porosity trades off with size in determining emergent Spectral Energy Distributions and determining Rosseland mean opacities. We use accurate material properties for a realistic, temperature-dependent, compositional suite, and simplified but realistic scattering and absorption
efficiencies which avoid the need for numerical Mie calculations. In this sense the approach is similar in spirit to some previous studies, but emphasizes a utilitarian approach and ease of general applicability. The physics
and simplifications are described in section 3 below; in section 4 we show some validation tests of the model, and in section 5 we describe the behavior of porous aggregates. Some basic derivations are presented in Appendices. The code itself is quite simple, taking negligible cpu time compared to a Mie code, and a version is available online. 

\section{Opacity model} 
In this section we describe the theoretical basis for our calculations of
particulate opacity. Our goal is to capture all of the significant physics in
the simplest fashion possible, so the model can be incorporated into evolutionary models at little computational cost. Realistic material refractive
indices are included for a cosmic abundance suite of likely nebula solids:
water ice, silicates, refractory organics, iron sulfide, and metallic iron. These refractive indices, and relative abundances as a function of temperature are taken from Pollack et al (1994; see Table). We chose these specific values to allow better validation and comparison with previous work, but they are widely used; alternate tabulations are easily incorporated, such as found in Draine and Lee (1984) or Henning et al (1999). 

\begin{table}
\begin{center}
\begin{tabular}{|c|c|c|c|c|c|c|c|}
\hline
 material & density & $\alpha_j$ & $T_{evap}$(K) & $\beta_j$ $<$160K  & $\beta_j$ $<$425K & $\beta_j$ $<$680K & $\beta_j$ $<$1500K \\
\hline
water ice  &0.9         &5.55e-3 & 160 & 6.11e-1 & 0 & 0 & 0\\
organics   &1.5        &4.13e-3 &425 & 2.73e-1 & 7.00e-1 & 0 & 0\\
troilite    &4.8       &7.68e-4 &680 & 1.58e-2 & 4.07e-2 & 1.36e-1 & 0 \\
silicates    &3.4      &3.35e-3 &1500 & 9.93e-2 & 2.55e-1& 8.51e-1 & 9.84e-1\\
iron ($<$680K)  &7.8  &1.26e-4 &1500 & 1.60e-3 & 4.11e-3 & 1.37e-2 & -\\
iron ($>$680K) &7.8    &6.15e-4 &1500 & - & - & 0 & 1.58e-2\\
\hline
\end{tabular}
\end{center}
\caption{Our assumed compositional mixture, adapted from Pollack et al (1994); for simplicity we have merged their two kinds of silicates and their two kinds of organics. When Troilite (FeS) decomposes at 680K, we follow Pollack et al in assuming the liberated iron adds to the existing iron metal and the S remains in the gas phase. The parameters $\alpha_j$ and $\beta_j$ define the fractional mass of a compositional species, and the fractional number of its particles if particles are segregated compositionally. For their definitions see section \ref{mixtures} and equations \ref{eq:betaj} (or \ref{eq:betaj1}) respectively.}
\end{table}

Our emphasis is on how size and porosity effects can be dealt with simply and seamlessly as grains aggregate and grow into a size regime where details of composition and refractive indices are perhaps secondary. Our basic
approach will be to separate particles having an arbitrary size distribution into two regimes which both have closed-form solutions in practical cases. We use the fact that particle interactions with radiation of any
wavelength $\lambda$ can be systematically addressed in terms of the
optical size of the particle $x = 2 \pi r / \lambda$, where $r$ is the
particle radius. The model presented here treats ``small" particles ($x \ll 1$)
as volume absorbers/scatterers and ``large" particles ($x \gg 1$) as
geometrical optics absorbers/scatterers.  We will occasionally refer to Van de Hulst (1957; henceforth VDH), Hansen and Travis (1974; henceforth HT), and Bohren and Huffman (1983). 

\subsection{Extinction, absorption, and scattering efficiencies}

Radiation is removed from any beam at a rate:
\begin{equation}\label{eq:1}
I = I_0 e^{- \kappa_e \rho l} 
\end{equation} 
where $\rho$ is the volume mass density of the gas and dust mixture, $\kappa_e$ (cm$^2$ g$^{-1}$) is the {\it total opacity}, 
and $l$ is the path length. The monochromatic opacity, for a gas-particle
mixture containing particles of radius $r$ with number density $n(r)$,
is formally defined as:
\begin{equation}\label{eq:2}
\kappa_{e,\lambda}= \frac{1}{\rho_g } \int \pi r^2 n(r) Q_e(r,\lambda) dr ,
\end{equation}
where $Q_e$ is the extinction efficiency for a particle of radius $r$, at wavelength $\lambda$. The Rosseland mean opacity, which controls the flow of energy through an optically thick medium, is derived by an appropriate weighting of $\kappa_{e,\lambda}$ over wavelength (see Appendix A). Extinction is due to a combination of pure absorption (reradiated as heat) and scattering
(redirection of the incident beam). Rigorously, these components are additive;
that is $Q_e = Q_a + Q_s$, where $Q_a$ is the absorption efficiency and $Q_s$
the scattering efficiency. The single-scattering albedo of the particle is
$\varpi  = Q_s/Q_e$. To the extent that the particles do scatter radiation
without absorbing it ($Q_s \ne 0$), the angular distribution
of the scattered component (the phase function $P(\Theta)$) is relevant.  The first moment
of the phase function $g$ describes the degree of forward scattering:
\begin{equation}
g =  \left< {\rm cos} \Theta \right> ={\int P(\Theta) {\rm cos}\Theta {\rm sin} \Theta d \Theta \over  
          \int P(\Theta) {\rm sin} \Theta d \Theta }
\end{equation}
Isotropic scattering results in $g=0$; very small particles (Rayleigh
scatterers) have a phase function proportional to cos$^2\Theta$, which also
leads to $g=0$. This is why scattering by tiny particles is often approximated
as isotropic. Large particles have a strong forward scattering lobe due to
diffraction; for such particles $g$ may approach unity ({\it cf.} HT) and can even be approximated as unscattered (Irvine 1975). In may cases of thermal emission when $r \ll \lambda$ (section \ref{abs}), only the absorption/emission component is important:
\begin{equation}\label{eq:abscoef}
\kappa_{a,\lambda}= \frac{1}{\rho_g } \int \pi r^2 n(r) Q_a(r,\lambda) dr ,
\end{equation}

Analytical expressions for $Q_a$ and $Q_s$ have been known for over a century
in certain asymptotic regimes; the well-known Rayleigh scattering regime 
for particles much smaller than the wavelength is one example. Similarly,
scattering properties of very large objects have long been well understood in
terms of Lambertian and related scattering laws. Considerable amounts of
computation have been devoted in recent years to determination of efficiencies
and phase functions in the intermediate Mie scattering regime, where particle size is
comparable to the wavelength. 

One important point for our purposes, that is
demonstrated by HT, is that many of the exotic fluctuations in scattering and
absorption properties which characterize the Mie regime vanish if one averages
over a broad distribution of particle sizes. Further smoothing occurs given a
combination of realistic particle nonsphericity and random particle orientations.
Furthermore, a large number of experimental studies (see also Bohren and Huffman 1983 and Pollack and Cuzzi 1980 for references) have shown that {\it integrated} parameters such as $Q$ and $g$ are much less affected by shape effects than, for example,
the phase function itself, and the parameters of interest for thermal emission and absorption ($Q_a, Q_s$, and $g$) vary smoothly between the ``Rayleigh" and ``geometric optics" regimes. This observed behavior provides a certain comfort level for the simple assumptions we make here. Our model does not treat $P(\Theta)$ at all, but constrains $g$ directly based on a fairly well-behaved dependence on the size and composition of the scatterer (section \ref{effs}). We use $g$ to correct $Q_a$ and $Q_s$ for energy which is primarily scattered forward, and thus does not participate significantly in directional redistribution of energy. Thus, our model would not be appropriate for applications where scattering dominates absorption {\it and} the directional distribution of scattered energy is of primary interest. Fortunately, primitive aggregate particles made of ice, silicates, carbon-rich organics and tiny metal grains are good absorbers and poor scatterers across most size ranges, but $Q_s$ does contribute to $\kappa_{\lambda}$ in some regimes.

\subsection{The model: Calculation of efficiencies}\label{effs}

A good exposition of the form of the particle absorption and scattering
coefficients is presented by Draine and Lee (1984; henceforth DL); our
expressions may be derived directly from those published by DL, and we will not repeat much of their presentation. DL describe
the radiative properties in terms of absorption and scattering cross sections
$C_{a,s} = Q_{a,s} \pi r^2$, which are directly related to particle volume in the limit $r \ll \lambda$. The
efficiencies may then be expressed in terms of the electric polarizability (VDH p. 73), which is a function of the (complex) particle dielectric constant $\epsilon =
\epsilon_1 + i \epsilon_2$ (DL eqn 3.11). The particle refractive index $m =
n_r + in_i$ is the square root of the dielectric constant. In general, $Q_a$ and $Q_s$ depend on the particle shape and orientation as well as the material refractive indices. The ``small dipole" limit, in which the particles are {\it both} highly nonspherical {\it and} have extremely large refractive
indices, has been discussed by DL, Wright (1987), Fabian et al (2001), and Min et al (2008). Lindsay et al (2013) show that even the shape of tiny solid grains can be diagnostic in high spectral resolution observations of strong absorption bands (section 2). After a small amount of algebra, DL equations 3.3, 3.6, and 3.11 lead directly to values of $Q_{ak}=C_{ak}/\pi r^2$ and $Q_{sk}=C_{sk}/\pi r^2$, where subscript $k$ refers to alternate orientations of a nonspherical particle relative to the wave vector:
\begin{equation}\label{eq:longerqa}
Q_{ak} =  {2 \over r^2 \lambda}{V \over L_k^2}
        { \epsilon_2 \over (L_k^{-1} + \epsilon_1 -1)^2 + \epsilon_2^2 }
     = \frac{4}{3L_k^2} \left( \frac{2 \pi r}{\lambda}\right){ \epsilon_2 \over (L_k^{-1} + \epsilon_1 -1)^2 + \epsilon_2^2 } 
\end{equation}
and
\begin{equation}\label{eq:longerqs}
Q_{sk} =  {8 \over 3r^2}\left({2 \pi \over 
                                          \lambda}\right)^4\left({V \over 4 \pi L_k}\right)^2
        { (\epsilon_1 - 1)^2 + \epsilon_2^2 \over 
                            (L_k^{-1} + \epsilon_1 -1)^2 + \epsilon_2^2 }
  = \frac{8}{27L_k^2}\left( \frac{2 \pi r}{\lambda}\right)^4 { (\epsilon_1 - 1)^2 + \epsilon_2^2 \over 
                            (L_k^{-1} + \epsilon_1 -1)^2 + \epsilon_2^2 }, 
\end{equation} 
where $V$ is the particle volume and
\begin{equation}\label{eq:dielconst}
\epsilon_1 = n_r^2 - n_i^2 ; \hspace {0.5 in} \epsilon_2 = 2 n_r n_i.
\end{equation}
Only the expression for $Q_a$ is explicitly given in DL (compare our equation \ref{eq:longerqa} with their equation 3.12). For the case at hand, however, once any significant amount of particle accumulation occurs, aggregates are likely to be roughly equidimensional and, especially if the particles are porous, the {\it average} refractive indices are not large (see below), so we will treat $Q_a$ and $Q_s$ as independent of particle orientation and, setting $L_k=1/3$, obtain 
\begin{equation}\label{eq:longqa}
Q_{a}     =  12\left(\frac{2 \pi r}{\lambda}\right){ \epsilon_2 \over ( \epsilon_1 +2)^2 + \epsilon_2^2 }
\end{equation}
and
\begin{equation}\label{eq:longqs}
Q_{s}  = \frac{8}{3} \left( \frac{2 \pi r}{\lambda}\right)^4 { (\epsilon_1 - 1)^2 + \epsilon_2^2 \over (\epsilon_1 +2)^2 + \epsilon_2^2 },
\end{equation} 
(VDH pp. 71, DL84; see also equation (14) of Miyake and Nakagawa 1993). Neither Miyake and Nakagawa (1993) or Pollack et al (1994) discuss $Q_s$, but it becomes important for porous, weakly absorbing particles in the transition regimes we wish to bridge with our model, where wavelength and particle size are not extremely different. In the geometrical optics regime, nonspherical shapes can increase the surface
area per unit mass, and thus the extinction efficiency, but for moderate nonsphericity the effect is only some tens of percent (Pollack and
Cuzzi 1980). This effect is also neglected for the present.  For plausible aggregates, magnetic effects are not important (see however Appendix B). 

In the limit where $n_i \ll n_r -1$, which is reasonable for ices, silicates,
organics, and their ensemble aggregates, equations \ref{eq:longqa} and \ref{eq:longqs} reduce to the simpler and more familiar forms:
\begin{equation}\label{eq:shortqa}
Q_a = { 24 x n_r n_i \over (n_r^2 + 2)^2}
\end{equation}
and
\begin{equation}\label{eq:shortqs}
Q_s = {8 x^4 \over 3} {(n_r^2 - 1)^2 \over (n_r^2 + 2)^2},
\end{equation} 
where $x=2\pi r/\lambda$. These are essentially the classical expressions given by VDH (p.
70). 

In our code we actually use the full equations \ref{eq:longqa} and \ref{eq:longqs} above for $Q_a$ and $Q_s$ for the radiative behavior of particles much smaller
than the geometric optics regime; these equations are valid for arbitrary
$m$ (DL). Equation \ref{eq:longqa} for $Q_a$ (linear in
particle radius) is used for all sizes. However, in the Mie transition region
(in which the particle size is {\it comparable} to a wavelength) we modeled
$Q_s$ using a function valid for $n_r$ of order unity (the Rayleigh-Gans regime; VDH p132-133, p.182). That is, at a transition value $x_0$ we change from the DL
expression (equation \ref{eq:longqs}) to the Rayleigh-Gans regime expression (VDH p. 182 and final equation of section 11.23):
\begin{equation}\label{eq:raygans}
Q_s = \frac{1}{2} (2x)^2(n_r - 1)^2 \left(1 + \left({n_i \over n_r -1}\right)^2\right),
\end{equation}
The transition value $x_0 = 1.3$ comes from setting the two expressions equal to each other.
This transition from $x^4$ to $x^2$ dependence bridges the region between the
Rayleigh and geometric optics regimes. It is shown in VDH (p 177) that a
coupled dependence on the so-called ``phase shift" $2x(n_r-1)$ (for $n_i << n_r
- 1$), which includes both $n$ and $r$,  captures the peak and shape of $Q_s$.   These expressions are valid as long as $n_r-1$ is not much larger than $1/x$ (Bohren and Huffman 1983, p.140); this is the expected regime for aggregates and mixtures of plausible materials. Our expressions do not capture the Fresnel-like oscillatory behavior of $Q$ shown by the the more complete series expansions in VDH and the full Mie theory, but do capture the correct small- and large-$x$ behavior, and the proper transition size/wavelength. This is an increasingly good
approach as the particle size distribution gets broader and the particles
become more absorbing (HT), in which case the resonance behavior near $2x(n_r-1)  \sim $ several diminishes and indeed vanishes (see, {\it eg.,} Cuzzi and Pollack 1978). Aggregate particles considered here are very good absorbers
in general, having very broad size distributions, but we show below that the technique works surprisingly well in even more challenging regimes.

\begin{figure}[t]                              
\centering                                                                   
\includegraphics[angle=0,width=3.1in,height=3.1in]{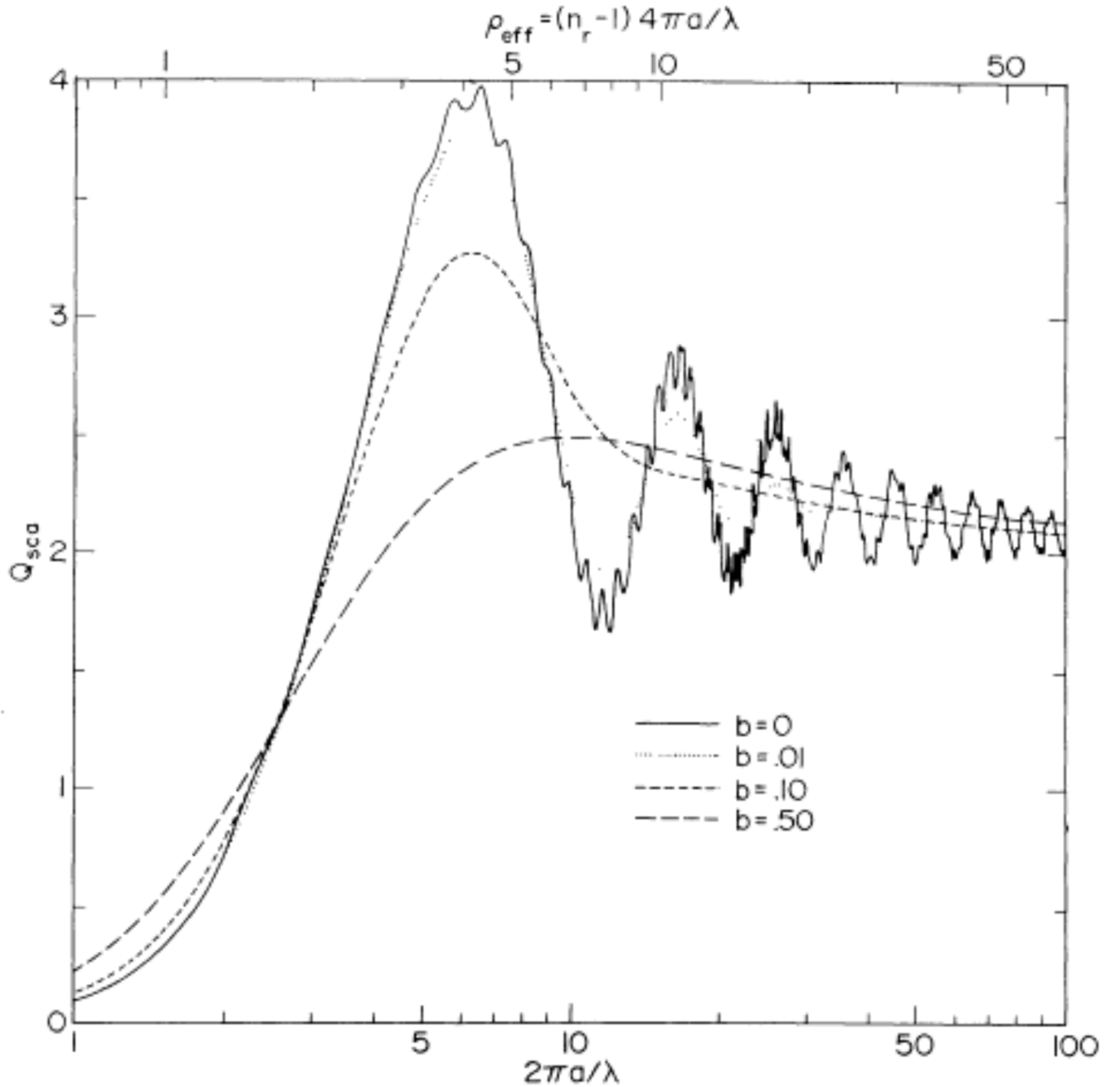}
\includegraphics[angle=0,width=3.1in,height=2.9in]{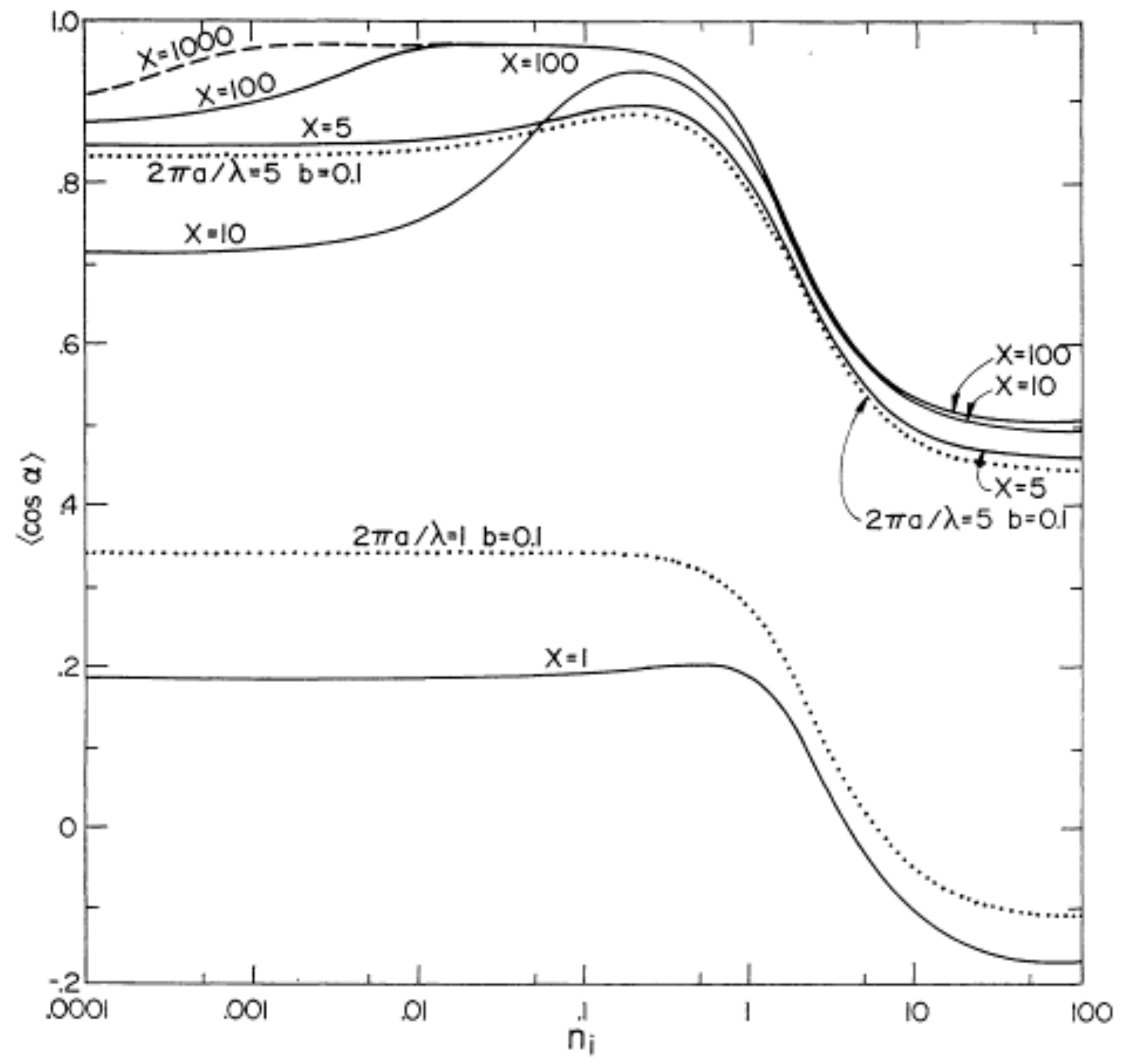}
\vspace{-0.2in}
\caption{Left: Scattering efficiency $Q_s$ of a particle with refractive indices $n_r=1.33,\,n_i=0$ as a function of its optical size $x=2 \pi a / \lambda$ using Mie Theory (figure reproduced directly from Hansen and Travis 1974, by permission, where particle radius = $a$). The parameter $b$ varies the width of the size distribution; as the size distribution gets broad, exotic ripples dues to interference average out and vanish. Right: dependence on the optical size and material properties of the scattering asymmetry parameter $g=\left< {\rm cos}\Theta \right>$, for narrow size distributions of different optical size. Note there are essentially two well-defined regimes in $n_i$. Figure \ref{fig:gvar} gives more detail on how $g$ varies with $x$.}
\label{fig:HT}  
\end{figure}

A key simplifying assumption of our model is that, while $Q_s$ and $Q_e$ grow from small values as $x$ increases, following equations (\ref{eq:longqa}) and (\ref{eq:longqs} or \ref{eq:raygans}),  {\it they are truncated at values representing the geometrical optics limit}, as discussed below. See for instance figure \ref{fig:HT} left, where in the geometric optics limit $2 \pi r/\lambda \rightarrow \infty$, $Q_e = Q_s \rightarrow 2$. The figure represents lossless particles; if absorption is present, $Q_s$ (which includes diffraction) trends downwards and asymptotes to a value closer to unity (HT figure 9). Note that the exotic ripples are primarily seen for monodispersions and narrow size distributions, and vanish as broader size distributions are used. 

Although the full radiative transfer problem must be solved when the {\it
angular variation} of the {\it intensity} is of primary concern, so that the phase function $P(\Theta)$ is important, certain
valuable simplifications are common when only angle-integrated quantities such
as {\it energy} or {\it flux} are of concern. In such cases, one often merely
corrects the extinction efficiency for the overall degree of forward- or
back-scattering. We will adopt a common scaling (e.g., Van de Hulst 1980) sometimes referred to as the radiation pressure scaling, and
adjust our efficiencies to take account of nonisotropic scattering as follows:
\begin{equation}\label{eq:qcorr}
Q'_e = Q_e - g Q_s = Q_e(1 - \varpi_0 g) = Q_a + Q_s(1 - g).
\end{equation}
That is, to the degree that scattered radiation is concentrated into the
forward direction, it may be regarded as unremoved from the beam. For instance,
even a large scattering efficiency contributes nothing to the extinction if
the scattering is purely forward directed ($g=1$). Negative values of $g$
(preferential backscattering) increase the extinction efficiency; in this
case, it is even harder for radiation to escape than for thermalized or
isotropically scattered radiation. This same correction is applied by Pollack et al (1985, 1994); see section \ref{por_eff} for more discussion. Modeling efforts relying to any degree on the Eddington approximation, in which scattered radiation is nearly isotropic, are perhaps better conducted using extinction efficiencies that are scaled in the way shown above, instead of using the formal Mie calculations (see Irvine 1975).  For instance, de Kok et al (2011) show how forward scattering by plausible exoplanet cloud particles can influence emergent near infrared fluxes; the above treatment of forward scattering partially accounts for this behavior by truncating away strong forward scattering and allowing the non-truncated scattered component to be treated as isotropic. In the approach presented here, the normal optical depth is defined using a value of $\kappa_{\lambda}$ that has been adjusted for forward scattering effects following equations \ref{eq:2} and \ref{eq:qcorr} (and \ref{eq:trunc} if appropriate): 
\begin{equation}
\tau_{\lambda}(z) = \int_z^{\infty} \kappa_{\lambda}(z) \rho_g(z) dz.
\end{equation}

\begin{figure}[t]                              
\centering                                                                   
\includegraphics[angle=0,width=3.1in,height=3.1in]{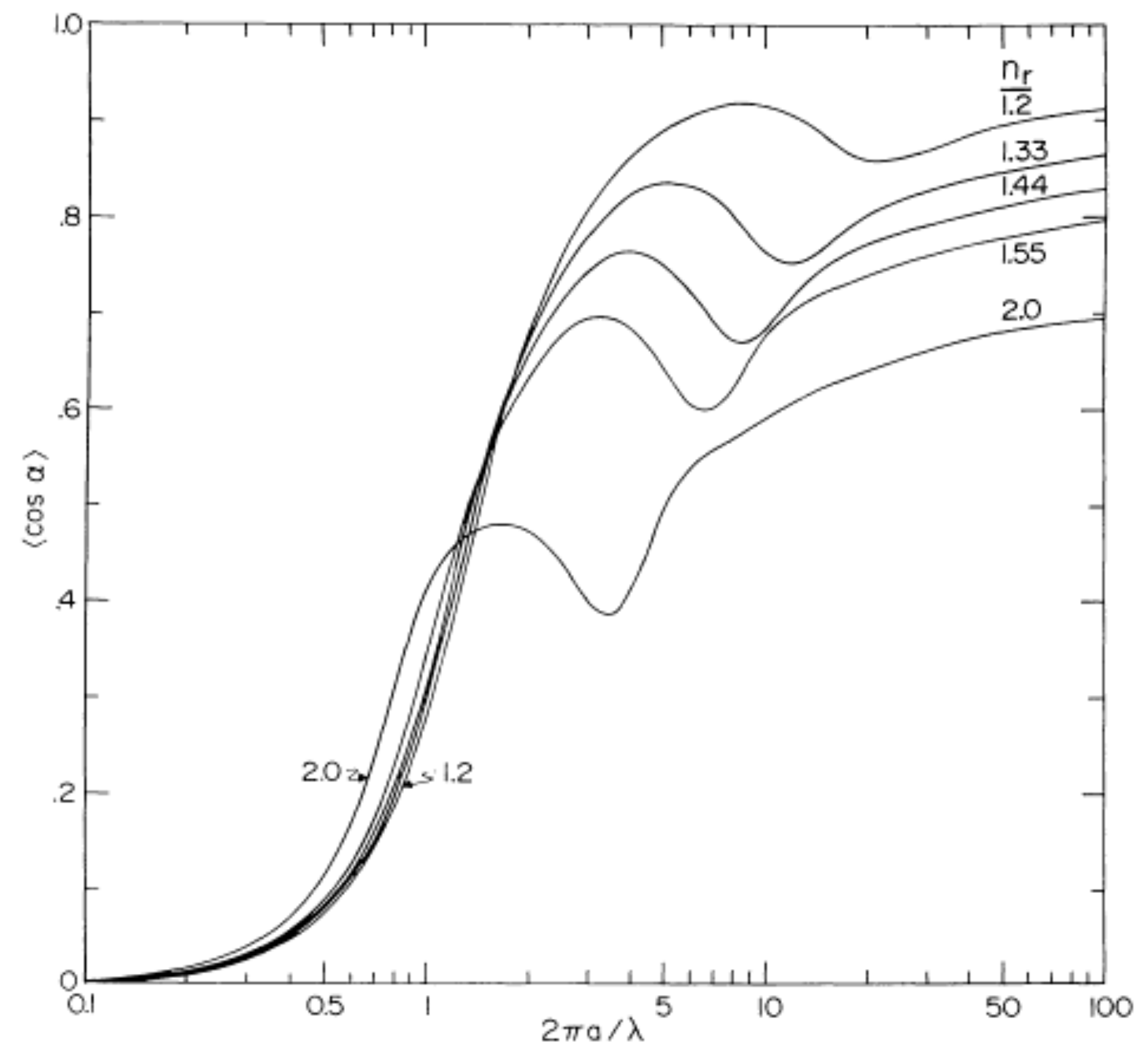}
\vspace{-0.2in}
\caption{Variation of $g = \left<{\rm cos}\Theta\right>$ ($\alpha = \Theta$ and $a=r$ in the notation of HT), and $x=2 \pi r / \lambda$, for a range of $n_r$. It is easy to show that the dependence in the steep transition region is $ g \propto x^2$, while the large-$x$ asymptote and transition point depend on $n_r$. With an eye towards the Garnett EMT value of $n_r$ for our cosmic mixture (1.7-2.1), we have chosen the single function $g=0.7(x/3)^2$ to represent all particles and materials. Figure reproduced from HT figure 12, by permission.}
\label{fig:gvar}  
\end{figure}

Instead of performing full Mie calculations to obtain the scattering asymmetry
parameter $g=\left< {\rm cos}\Theta \right>$, we make use of the crudely partitioned behavior illustrated by
HT (reproduced in figure \ref{fig:HT}  (right)). We have explored several simple ways of doing this, and have chosen the following determination of $g$, consistent with figures \ref{fig:HT} and \ref{fig:gvar} ( notice that figure \ref{fig:HT}(right) is for a fairly low $n_r=1.33$):  
\begin{equation}\label{eq:g3}
{\rm for}\,\, n_i < 1: g=0.7(x/3)^2 \hspace {0.1 in} {\rm if} \hspace {0.1 in} x < 3, \,\,{\rm and}\,\, 
g \approx 0.7 \hspace {0.1 in} {\rm if} \hspace {0.1 in} x > 3,
\end{equation}
\begin{equation}\label{eq:g4}
{\rm for}\,\, n_i > 1: g \approx -0.2 \hspace {0.1 in} {\rm if} \hspace {0.1 in} x < 3, \,\,{\rm and}\,\, 
g \approx 0.5 \hspace {0.1 in} {\rm if} \hspace {0.1 in} x > 3.
\end{equation}
The value of the $g$-asymptote at large $x$, and the transition value of $x$, vary  slightly with $n_r$ (figure \ref{fig:gvar}); the values we selected apply to the EMT values of $n_r \sim 1.7-2.1$ and $n_i \ll n_r$, most relevant to our mixed-aggregate applications. We have not attempted to fine-tune either the asymptote or the transition value, and leave the large-$n_i$ definition of $g$ in its crude bimodal form (the large-$n_i$ recipe has almost no effect on the results). Further fine-tuning is somewhat beyond what we expect of a simple utilitarian model, but could be done easily.

The final piece of the model involves matching the growing values of $Q_s$ and $Q_a$ in the Rayleigh and/or Rayleigh-Gans regime to constant values in the geometric optics regime, simply by limiting their magnitudes:
\begin{equation}\label{eq:trunc}
Q_a < 1 \hspace{0.2in} {\rm and} \hspace{0.2in} Q_s(1-g) < 1.
\end{equation}
This can be thought of as limiting the absorption cross section, and the diffraction-corrected scattering cross section, to the physical cross section. 
D'Alessio et al (2001) did not adjust their $Q$, albedo, or $\kappa$ values for $g$ as do we and Pollack et al (1994), instead taking them straight from their Mie code. If one then carries on to solve the radiative transfer equations in an Eddington-like approximation (that is, as in their equation (3), assuming isotropic scattering, or otherwise without detailed treatment of the angular dependence of often strongly forward-scattered radiation), indeed it would be better to first perform the truncation following equation \ref{eq:qcorr}. A complete multidimensional treatment can capture all this behavior of course. D'Alessio et al (2006) compared forward scattering with isotropic scattering, showing very little difference (their figure 7)\footnote{Some of the curves in the lower left panel of Dalessio et al 2006 seem to be mislabeled, but the trends are fairly clear.}, but at their longer wavelengths $r \ll \lambda$ so scattering is negligible in the first place, and at shorter wavelengths the emission arises from the photosphere of an opaque disk. A detailed discussion of this issue is beyond the scope of this paper. 

One final detail must be mentioned, relevant for applications where particles of pure metal are important. Equations (\ref{eq:longqa}) and (\ref{eq:longqs} or \ref{eq:raygans}) include only the electric dipole interaction terms that are appropriate for everything but metals, and thus for all of our aggregate particle models where refractive indices are not extremely large. However, when we compare our model with the heterogeneous-composition grain mixture of Pollack et al (1994), in which pure iron metal grains play a role, we have used a crude correction to $Q_a$ for magnetic dipole terms as well, following DL84, Pollack et al (1994), and others (see Appendix B); these calculations are all compared with the full Mie theory below. 

\begin{figure*}[t]                             
\centering                                                                   
\includegraphics[angle=-90,width=6.2in,totalheight=4.0in]{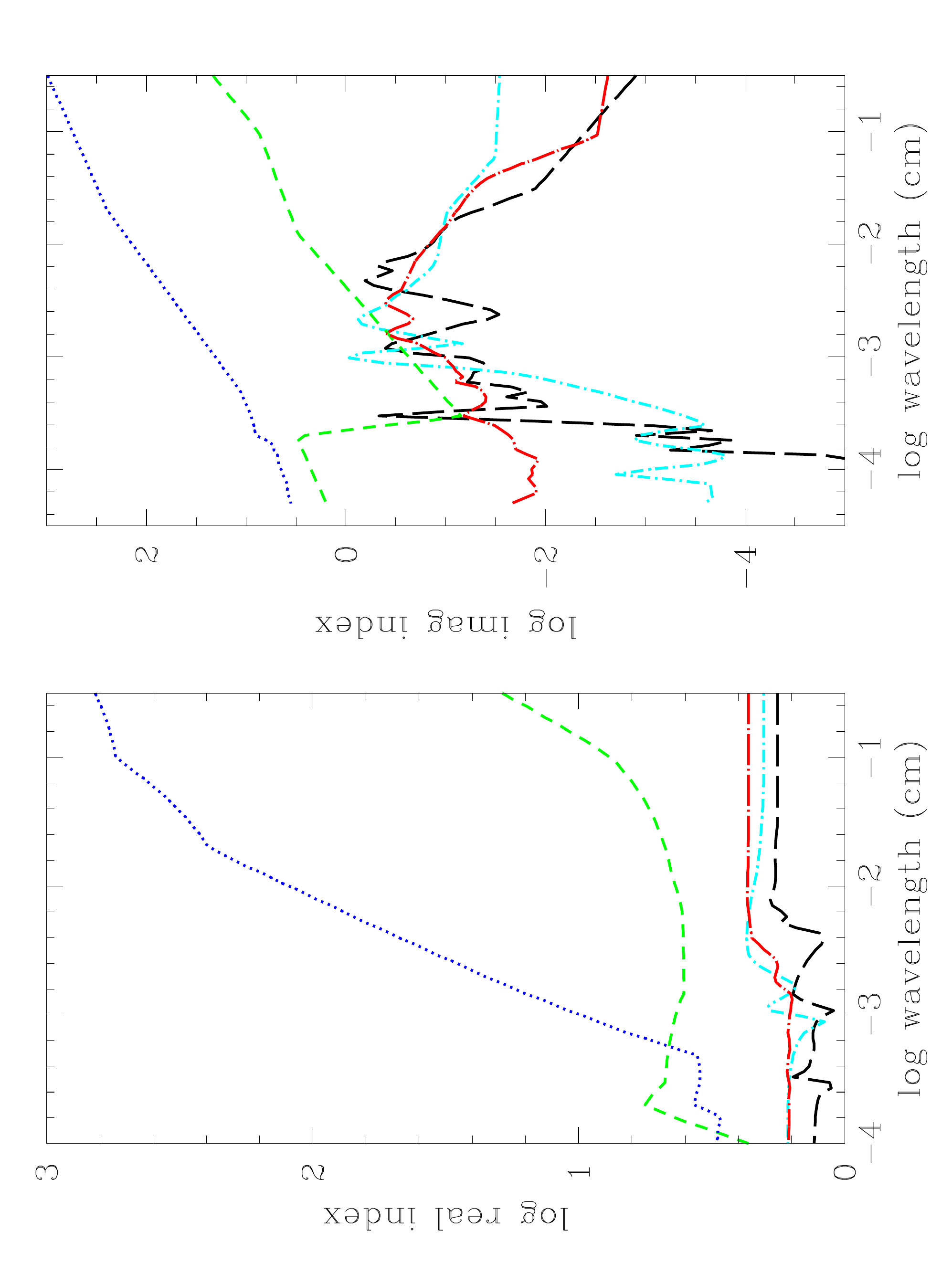}
\vspace{-0in}
\caption{Real (left) and imaginary (right) indices as used in our code, representing a number of common materials, taken directly from Pollack et al (1994). Black long dash:  water ice; Cyan dash-dot: silicates; Green short dash: Iron Sulfide; Blue dotted: Iron metal; Red long dash-dot: organics. For simplicity we have used only one of Pollack et al's two varieties of silicates (pyroxene) as they both have similar properties, and only one variety of organics (the two of Pollack et al (1994) had the same refractive indices but two different evaporation temperatures were assumed).}
\label{fig:indices}      
\end{figure*}

\subsection{Material Properties}
For ease of comparison with previously published results, we adopt refractive indices, material abundances, and stability regimes for
the condensible constituents as published by Pollack et al
(1994). We have merged the two different silicates (olivine and pyroxene) and the two different organics (higher and lower volatility) of Pollack et al (1994) into a single population of each for simplicity.  The exact composition of protoplanetary nebulae is, after all, poorly known, but these are consistent with known properties of grains in primitive
meteorites and comets and cover an appropriate range of evaporation temperature.  The components, their assumed relative abundances, and their evaporation temperatures are shown in the Table. We will show that particle growth leads to at best a weak dependence on detailed stipulations of composition.  It would be simple for anyone to select an alternate mixture. 

Our refractive indices are shown in figure \ref{fig:indices}. The materials can be classified into two groups - the water ice, silicate, and organic material have real refractive index $n_r$ on the order of unity, and
imaginary refractive index $n_i$ much less than unity. On the other hand,
metallic iron and iron sulfide particles have both refractive indices larger (often much larger) than unity. 

\section{Mixtures of grains with pure composition}\label{mixtures}

In our primary intended application to protoplanetary nebulae, particles are porous composites or granular mixtures of much smaller grains of all compositions that are solid at local temperature $T$, and we will calculate their ensemble refractive indices $(n_r,n_i)$ using the Garnett Effective Medium Theory (EMT) as described in Appendix C and section \ref{agg_ind} below. However, in some applications, such as exoplanet atmospheres, grains of a particular composition - even perhaps grains of iron metal - might be isolated within their own cloud layer. 

To assess our approach in this regime, we first calculate opacities for a compositionally {\it heterogeneous} 
ensemble of micron-size grains adopted by Pollack et al (1985, 1994) in their
exact Mie scattering calculations.  That is, distinct
populations of non-porous grains  with distinct compositions and refractive indices are assumed, and the opacities so determined are averaged as described below. This is a more challenging regime than a situation in which aggregates are of mixed composition and thus do not have extreme values of refractive index (section  \ref{agg_ind}).  
   
For particles of species $j$, the monochromatic opacity
at wavelength $\lambda$ (relative to the gas mass density $\rho_g$) is:
\begin{equation}\label{eq:kap_ej}
\kappa_{e,j,\lambda} = {\beta_j \over \rho_g} \int n(r) \pi r^2 
                       Q'_{e,j}(r, \lambda) dr
\end{equation}
where $n(r) = \sum n_j(r)$ and $n_j(r) = \beta_j n(r)$; $\beta_j$ is the fraction by number of particles of species $j$; that is, as assumed by Pollack et al (1994) for this particular case, all 
species are assumed to have the same functional form for their size 
distribution. The Pollack size distribution used is a differential powerlaw for the particle number volume density in some radius bin $dn(r,r+dr)=n(r)dr$ where $n(r)= n_o r^{-p}$, $p=-3.5$ for 0.05$\mu$m$<r<$1$\mu$m, $p=-5.5$ for 1$\mu$m$<r<$5$\mu$m, and zero otherwise. Note that this size distribution is actually quite narrow: it covers roughly one decade in radius nominally (the piece from 1-5$\mu$m contains very little mass or area), and is sufficiently steep for most of the surface area to be provided by the smallest members. In Appendix C we show that 
\begin{equation}\label{eq:betaj1}
\beta_j = { \alpha_j / \rho_j  \over  \sum (\alpha_j / \rho_j)}. 
\end{equation}
where $\alpha_j$ is the fractional {\it mass} of constituent $j$ relative to the gas,
and $\rho_j$ is the density of solid constituent $j$. Note that $\beta_j$ is the fractional {\it number} of particles of species $j$, and is subtly different in general from the fractional {\it volume} weighting factor $f_j$ that is used when compositions are mixed together in the same particle as in our EMT approach (for compact or non-porous particles they are the same). The above definition of $\beta_j$ differs slightly from that given in Pollack et al (1985); their equations (2)-(4) capture the right essence of the solution, but are not rigorously correct as written. By stepping back and incorporating the $Q$ factors (including the correction for scattering asymmetry) inside the integrals of their equation (4), the same result can be obtained: 
\begin{equation}\label{eq:heterosum}
\kappa_{e,\lambda} = \frac{1}{\rho_g} \sum_j \beta_j 
                     \int n(r) \pi r^2 Q'_{e,j}(r, \lambda) dr
                   = {1 \over \rho_g} \int n(r) \pi r^2 
                     (\sum \beta_j Q'_{e,j}(r, \lambda) ) dr,
                     \end{equation}
where $Q'_{e,j}$ includes the forward scattering correction (equation \ref{eq:qcorr}). This is probably the most challenging test we could pose - an ensemble of
heterogeneous particles, including some with very large refractive indices.
Some are good absorbers and some are very good and rather anisotropic
scatterers. 

\begin{figure*}[t]                               
\centering                                                                   
\includegraphics[angle=-90,width=0.9\textwidth]{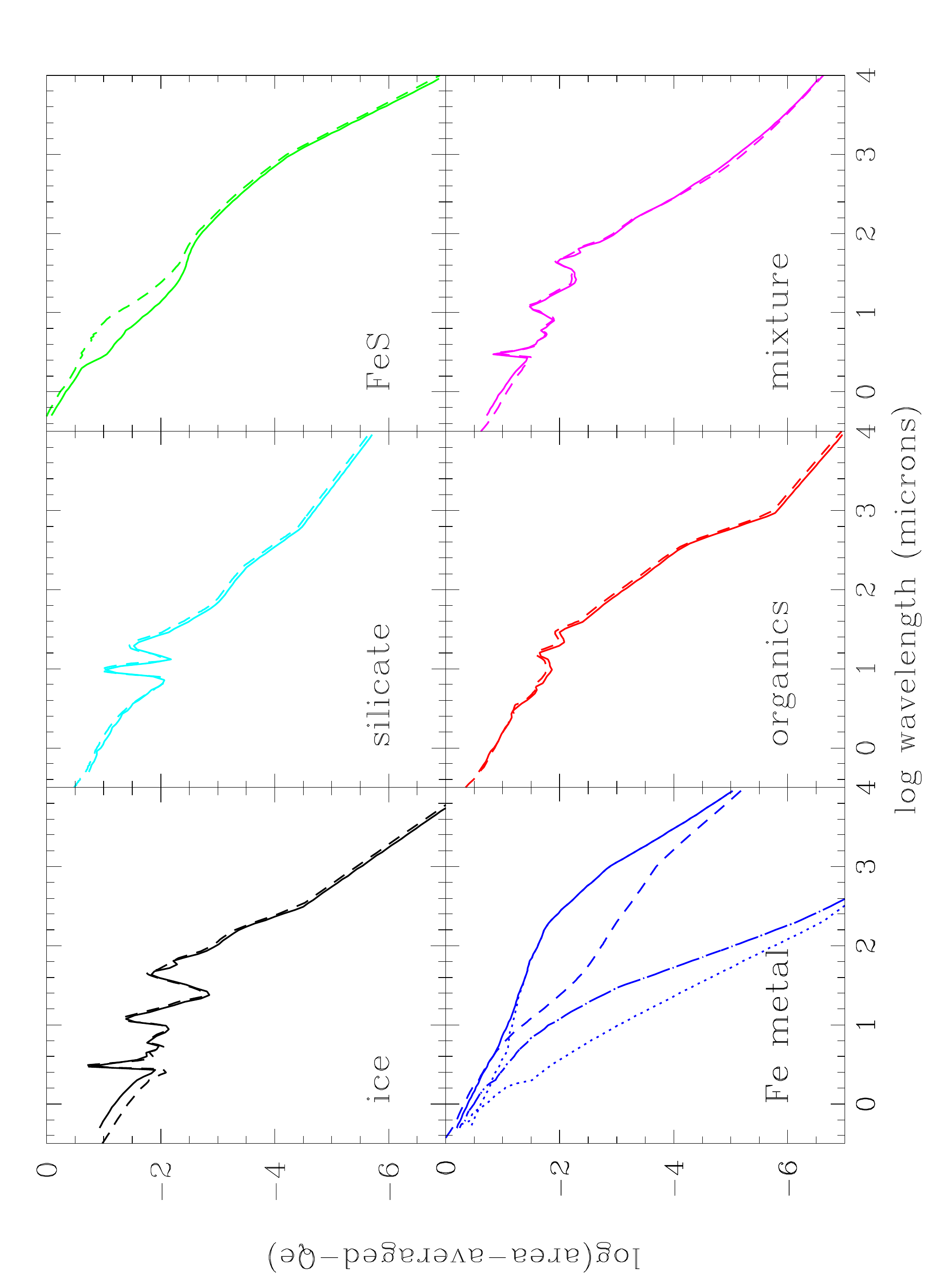}
\vspace{0in}
\caption{A comparison with Mie theory of our mean extinction efficiency $Q'_{ej}(\lambda)$ for five pure materials, averaged and weighted over the Pollack et al (1994) size distribution (see section 4.1). Our model results are shown in solid curves and Mie calculations are shown in dashed curves. The lower right panel is the abundance-weighted average value $Q'_e(\lambda)$. The solid curve in the iron panel includes an (imperfect) correction for magnetic dipole effects (Appendix B), and the dotted ($Q_a$) and dot-dash ($Q_s$) curves do not. Actually, all the solid curves incorporate this correction term, but it makes a difference only for iron. } 
\label{fig:6panela}  
\end{figure*}

\subsection{Wavelength-dependent opacities}
The fundamental calculation is of the effective, forward-scattering-corrected (or ``radiation") extinction efficiency for each compositional type $Q'_{ej}(r,\lambda)$ (equation \ref{eq:qcorr}, using equations \ref{eq:longqa} and \ref{eq:longqs} or \ref{eq:raygans}). If the situation calls for a heterogeneous mixture of grains with distinct compositions, $Q'_{ej}(r,\lambda)$ must be calculated separately for each composition. 

We have compared our model calculations directly with Mie calculations for five separate compositions, in each case averaged over the Pollack et al (1994) particle size distribution (which is fairly narrow) as weighted by particle number density and geometric area. That is: 
\begin{equation}\label{eq:weighted}
Q'_{ej}(\lambda) = { \int n(r)\pi r^2 Q'_{ej}(r,\lambda) dr \over
                    \int n(r)\pi r^2 dr}.
\end{equation}
These results are shown in figure \ref{fig:6panela}. The model does surprisingly well overall, except for metallic particles (FeS has a near-metallic behavior at short wavelengths). Note the different model curves in the iron metal panel; the solid curve attempts to correct for the effects of magnetic dipole interactions with the DL84 approximation also used by Pollack et al (1994) (section \ref{effs}). This term is only the leading term in an expansion, here used outside its formal realm of validity (see Appendix B). We would not advocate use of our simple model in cases where pure metal particle clouds might be encountered, such as in some exoplanet atmospheres (Lodders and Fegley 2002, Marley et al 2013). However, it does improve agreement (especially at long wavelengths) with the full Mie calculations of similarly size-and-abundance averaged values of both $Q'_e(\lambda)$ (bottom right panel of figure \ref{fig:6panela}) and $\kappa_R$ (figure \ref{fig:ross_comps} below), for the Pollack et al (1994) heterogeneous grain mixtures.  In fact, {\it all} panels in figure \ref{fig:6panela} include calculations done with and without magnetic dipole terms, but the differences are insignificant for all materials except iron metal (there is a small effect for FeS), and in the averaged values (lower right panel). These individual compositional efficiencies and/or cross-sections are then combined into an overall opacity (equation \ref{eq:heterosum}), or simply an average efficiency, as weighted by their abundances:
\begin{equation}\label{eq:weighted}
Q'_{e}(\lambda) = \sum \beta_j Q'_{ej}(\lambda). 
\end{equation}
A plot of $Q'_{e}(\lambda)$ is shown in the bottom right panel of figure \ref{fig:6panela}, also compared with the similarly summed Mie-derived values.  

The more plausible case for protoplanetary nebulae, where all the particles are single (mixed) composition aggregates, with refractive index given by the Garnett EMT (see Appendix C and section \ref{aggs} below), is even simpler. A single efficiency $Q'_{e}(r,\lambda)$, associated with the EMT refractive indices, is simply integrated over the size distribution to give some wavelength-dependent opacity $Q_e(\lambda)$. Whether determined from a compositionally heterogeneous or homogeneous mixture, $Q'_e(\lambda)$ is the basis of {\it extinction} calculations such as Rosseland opacities (equations \ref{eq:1} and \ref{eq:2}). 

\begin{figure*}[t]                                 
\centering                                                                   \includegraphics[angle=-90,width=3.1in,totalheight=2.5in]{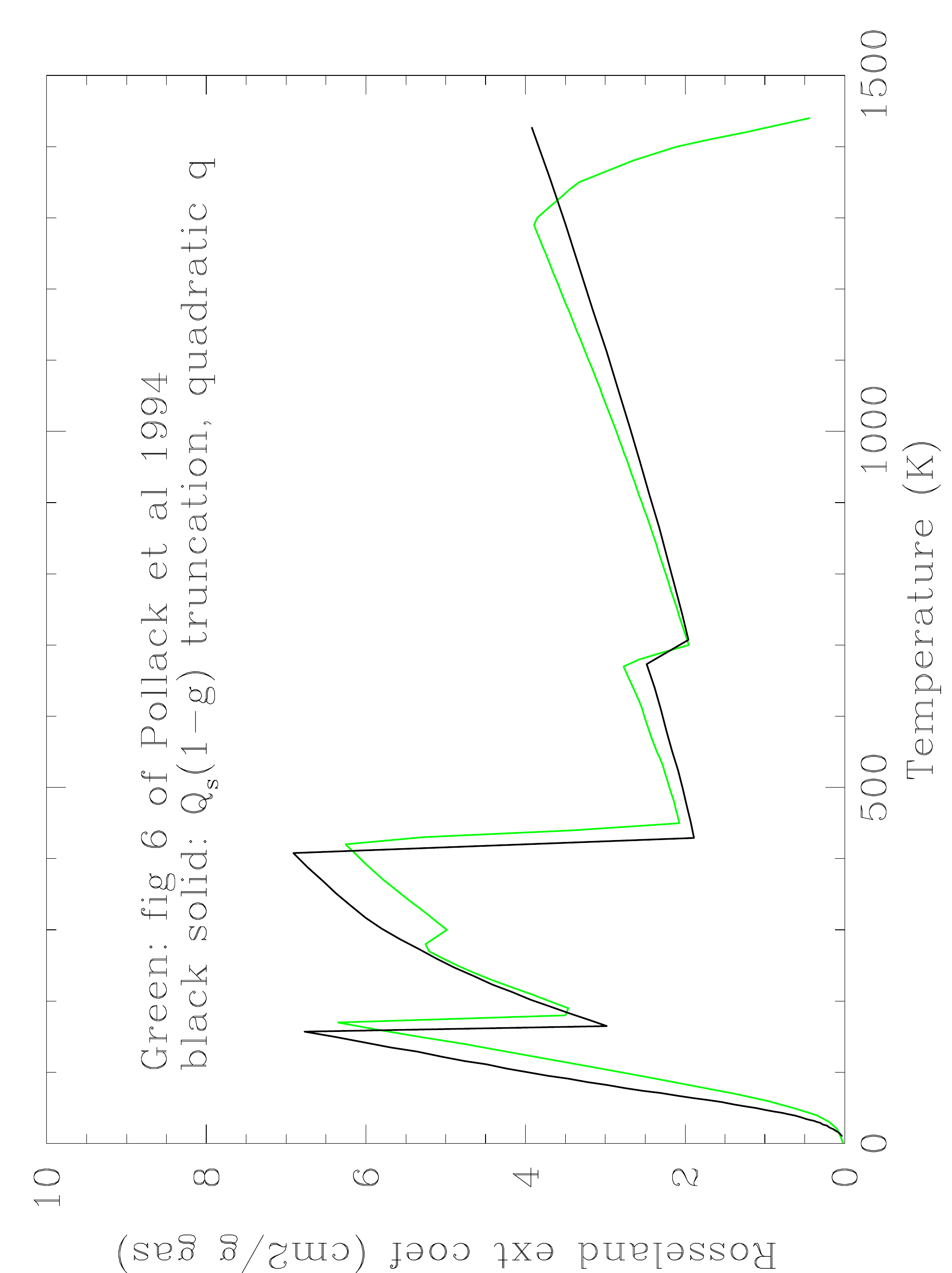}
\vspace{-0in}
\caption{A comparison between the Rosseland mean opacities as determined here, with much more elaborate full Mie scattering calculations (see text for discussion). The Pollack model has silicates, organics, and water ice evaporating at slightly different temperatures than assumed here (in fact they are a function of pressure); these are easily adjusted if desired. } 
\label{fig:ross_comps}
\end{figure*}

\subsection{Rosseland Mean opacities}
We initially apply our models to calculate Rosseland mean opacities $\kappa_R$ (Appendix A) for the standard Pollack et al (1994) heterogeneous grain size distribution. Figure \ref{fig:ross_comps} shows that our results agree quite well with the exact Mie scattering calculations of Pollack et al (1994), especially considering the simplicity of our approach.  The Pollack et al size distribution is an ISM distribution, with a steep powerlaw $n(r)$ between radii of .005 - 1.0$\mu$m and a steeper powerlaw from 1.0-5.0$\mu$m (section \ref{mixtures} above). Pollack et al (1994) introduce two types of moderately refractory organics (with the same refractive indices but different abundances and evaporation temperatures) to provide greatly increased opacity between 160K and 425K, which we combine into a single component; thus we lack one evaporation boundary at about 260K; a second difference is that we have only used a single silicate component. In view of these differences in detail, and for a situation like this (five distinct species including iron metal, with grain sizes in the resonance regime at the shorter wavelengths at high temperatures) which is far more challenging than our primary intended application (well-mixed aggregates of moderate refractive index, many in the geometrical optics limit), the agreement is surprisingly good. 

\section{Aggregate particles, each of mixed composition}\label{aggs}
In the remainder of the paper, we explore the radiative properties of more mature and (at least for protoplanetary nebula applications) probably more realistic particle porosity and size distributions between a micron and arbitrarily large sizes.  Evidence
from meteorite and interplanetary dust particle samples indicates that
accumulated particles of these larger sizes are heterogeneous aggregates of all candidate materials, with individual submicron-to-micron size ``elements" being composed of one mineral or other. Even mm-size chondrules are each a mixture of 1-10$\mu$m size mineral grains of silicate, iron, and iron sulfide, and the ubiquitous meteorite matrix is a mix of smaller grains of all these materials, perhaps originally aggregates. So, for our aggregates, we assume only one (mixed) grain composition; the assumption may be shaky at the very smallest sizes which may be mono-mineralic, but  coagulation models show that the tiniest particles are quickly consumed by growing aggregates (Ormel and Okuzumi 2013). Then, we are back to equation \ref{eq:2}:
\begin{equation}\label{eq:homog}
\kappa_{e,\lambda} = \frac{1}{\rho_g} \int n_0(r)\pi r^2 Q'_e(r,\lambda)dr.
\end{equation}

As the gas density (and grain number density) increases, the
collision rate increases accordingly. Grains of micron size are well coupled
to the gas, and have fairly low collision velocities relative to each other
(most recently Ormel and Cuzzi 2007). The low relative
velocities imply that sticking will be fairly efficient at least until
mm-cm sizes are reached (Dominik et al 2007, G{\" u}ttler et al 2010, Zsom et al 2010, Birnstiel et al 2010,2012). Several laboratory
simulations of this process have shown that the resulting aggregates are of
fairly low density (Donn 1990, Beckwith et al 2000). It has been noted that
aggregates of this sort are fractals in the sense that their volume depends on
their mass to an arbitrary power. Normally this relationship is expressed as 
$ m \propto r^s$, where $m$ is an individual particle mass and $s$ is the 
fractal dimension (Wright 1987, Beckwith et al 2000, Dominik et al 2007).
If the dimension is less than 3, their internal density decreases as the
mass increases. Experimental results suggest that the internal
densities and volume fractions of such particles are likely to be quite small, with porosities approaching 70\% even after they begin compacting each other at increasing relative velocities (Weidenschilling 1997, Ormel et al 2008 and references therein). Growth of this sort is certain to be robust in giant planet atmospheres as well (Ackerman and Marley 2001, Helling et al 2008, Marley et al 2013 and references therein). 

\subsection{Refractive indices of aggregate particles}\label{agg_ind}
\begin{figure}[t]                              
\centering                                                                   
\includegraphics[angle=0,width=3.1in,height=3.6in]{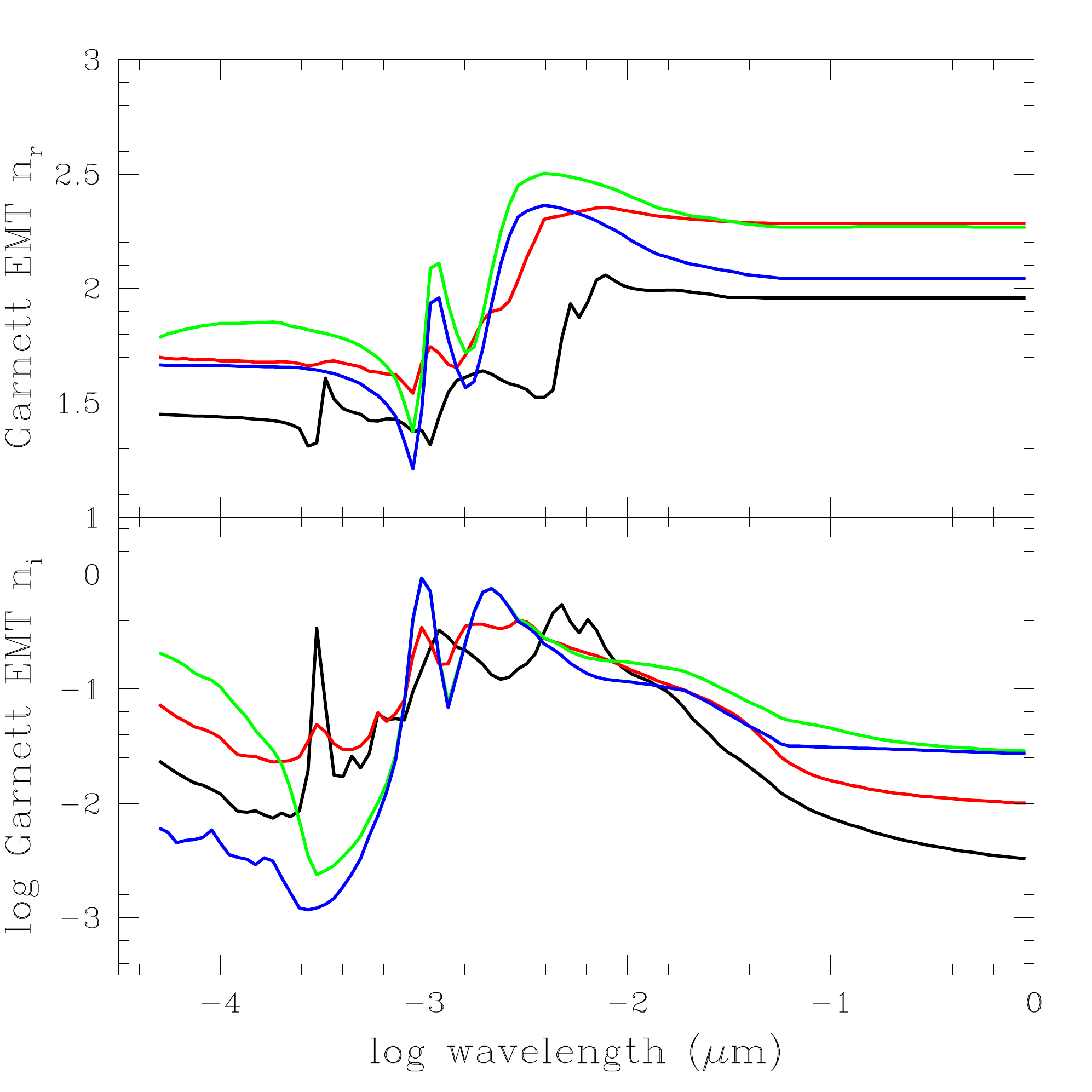}
\includegraphics[angle=0,width=3.1in,height=3.6in]{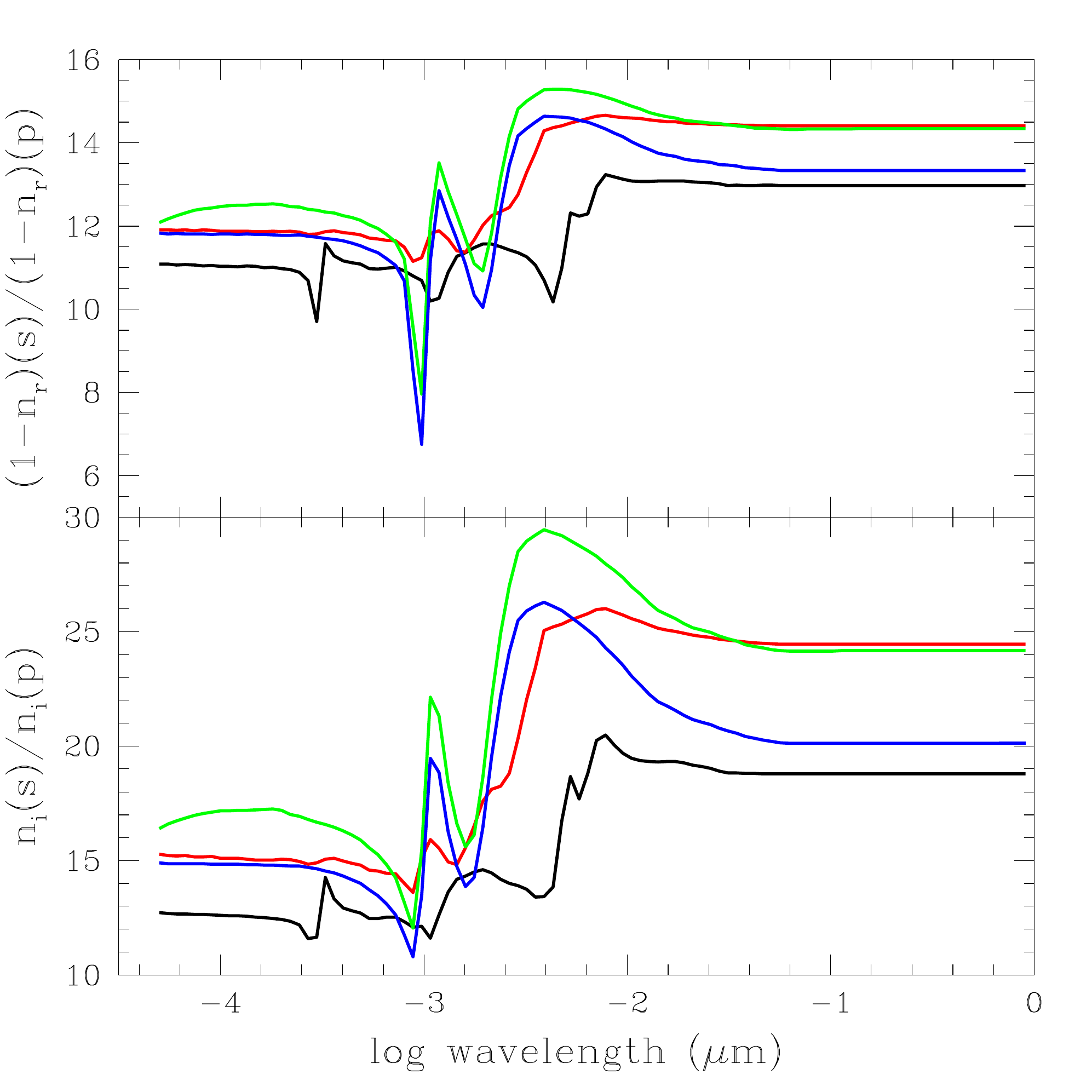}
\vspace{-0.1in}
\caption{Left: Refractive indices of aggregate particles, obtained using the Garnett EMT, as functions of wavelength for four different temperature regimes in which all materials are present (black), water ice has evaporated (red), organics have evporated (green) and troilite has evaporated (blue). The temperatures at which these transitions occur are shown in the Table. Right: ratio of solid/porous particle EMT refractive indices in the four temperature regions, for a porosity of 90\%. For porosities in this range, these refractive indices could simply be used in our basic equations \ref{eq:longqa} and (\ref{eq:longqs} or  \ref{eq:raygans}), along with equation \ref{eq:qcorr} as constrained by equations \ref{eq:g3}-\ref{eq:trunc}. These results would be used in our model to calculate particle opacities at temperatures where one or more of the five basic constituents has evaporated or decomposed (this is, of course, implicit in calculation of Rosseland mean opacities as a function of temperature, but the monochromatic opacities shown here are for a temperature where all candidate solids are condensed). Tabulated values available from the authors and will be posted online. }
\label{fig:mgindices}  
\end{figure}

We envision {\it each particle} as compositionally heterogeneous - made up of
much smaller constituents of specific composition and/or mineralogy,
as seen in meteorites, interplanetary dust particles (IDP's), and cometary
dust. That is, our particles are aggregates of subelements of species $j$, each smaller (and usually much smaller) than any relevant wavelength. The average particle refractive indices can then be calculated using an Effective Medium Theory (EMT) approach. The two best known EMTs are due to Garnett, and Bruggeman (Bohren and Huffman 1983). In the Garnett model, there is presumed to be one pervasive ``matrix" in which distinct grains of other materials are embedded. In the Bruggeman theory, there is no structural distinction between domains of different refractive index. Bohren and Huffman (1983) believe that the Garnett rule is fundamentally to be preferred for aggregate particles where there {\it is} a well-defined matrix and it is vacuum (even, in principle, as the porosity gets very small). As this is our application, we use the Garnett EMT. The expressions and some additional discussion are given in Appendix C, and a set of effective refractive indices for each temperature range (each ensemble of condensed solids) is shown in figure \ref{fig:mgindices}. 

In the 
limit where all of the refractive indices are of order unity (this holds 
away from absorption bands for all species but the iron and troilite), an 
intuitively simple linear volume average captures the sense of the effect and 
might serve acceptably well in some regimes:
\begin{equation}
n_i = \frac{1}{f}\sum_j  f_j n_{ij}
\end{equation}
and
\begin{equation}
n_r = 1 + \frac{1}{f}\sum_j  f_j (n_{rj} - 1).
\end{equation}
Here, $f$ is the {\it volume} fraction of all solids in a given particle, and $f_j$ is the volume fraction for each species $j$. For low mass
density particles such as some of those seen in our model distributions, the average imaginary index can get quite small and the real index can
approach unity. However, in  Appendix C and section \ref{wave_dependent} below, we demonstrate that such a simple volume average greatly overestimates the contribution of even a small volume fraction of high-refractive-index grains ({\it ie.,} iron and troilite) in the mixture; for the purpose of this paper only the full Garnett theory (described in detail in  Appendix C) is used.

\begin{figure*}[t]                                 
\centering                                                                   
\includegraphics[angle=0,width=3.3in,totalheight=3.3in]{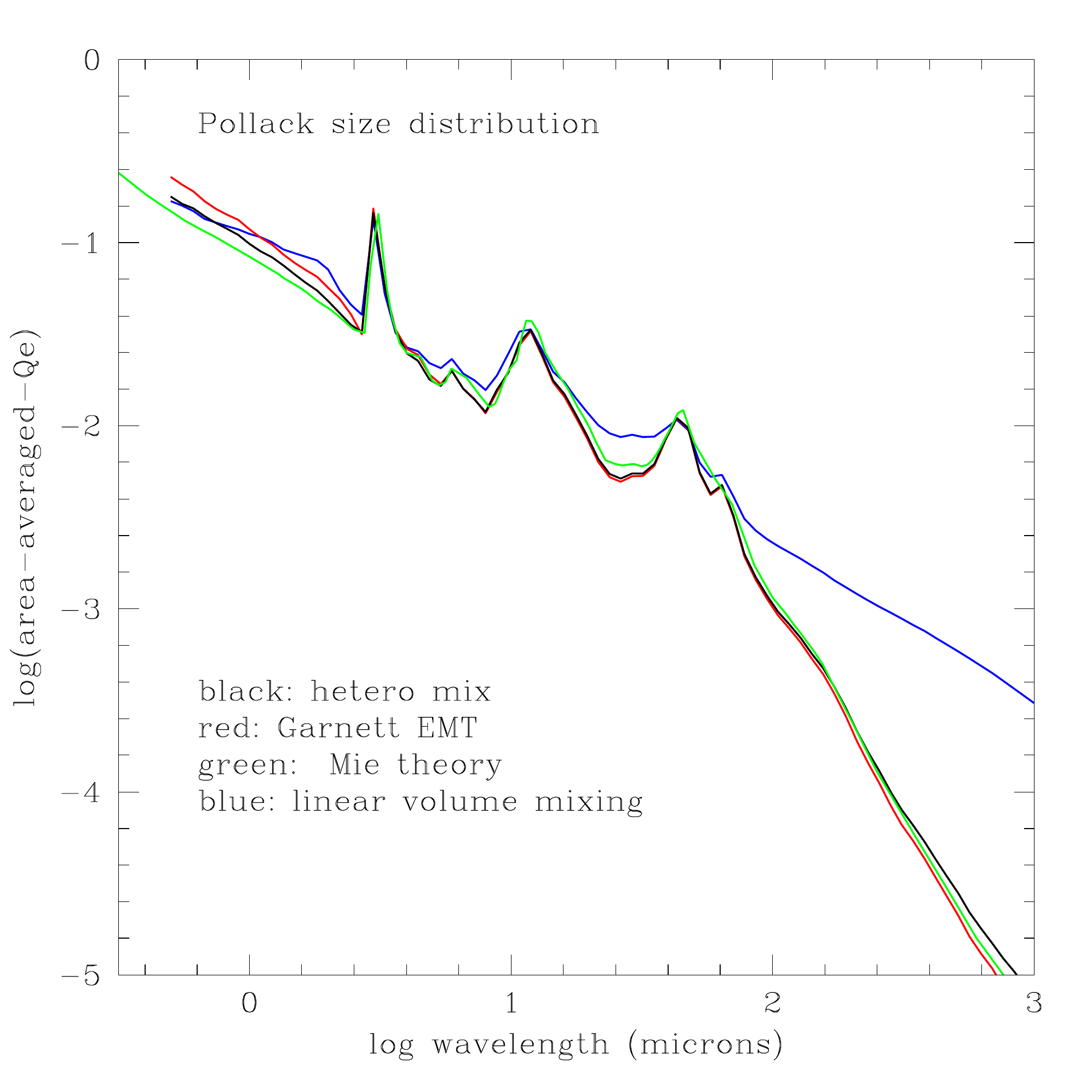}
\vspace{-0in}
\caption{A comparison of the wavelength-dependent mean extinction efficiency $Q'_e(\lambda)$, weighted and summed over the Pollack et al (1994) particle size distribution, for two particle structure assumptions. The heterogeneous mixture in which each composition $j$ is treated as a separate pure material with the same size distribution, and the various $Q'_{ej}(\lambda)$ then combined using their appropriate abundances, is shown in  black for our model and green for Mie theory. Also shown (red) is the Garnett EMT treatment of the same size distribution, but with each particle an identical aggregate of the five materials, along with (blue) the linear mixture ``approximation" to EMT. Note how badly the linear approximation performs, as also discussed in the Appendix. On the other hand, for these small particles, the heterogeneous mixture and the Garnett EMT results are very similar.} 
\label{fig:mix_garn}
\end{figure*}

\subsection{Wavelength dependent opacities}\label{wave_dependent}
In figure \ref{fig:mix_garn} we compare weighted averages of $Q'_e(\lambda)$ over heterogeneous mixtures, as described in section \ref{mixtures}, with values obtained for the same size distribution of aggregates in which the same material is mixed within each particle, and refractive indices are determined using the Garnett EMT (section \ref{agg_ind} and Appendix C). As expected for particles much smaller than the wavelength (section \ref{effs}), and as found by other investigators in the past, there is very little difference in the wavelength-dependent efficiency between a calculation in which each material is treated separately, and one in which they are merged together and treated with the EMT. The Pollack size distribution, while formally extending to 5$\mu$m radius, is very steep and in reality has nearly all the cross section and mass at radii smaller than, if not much smaller than, 1 $\mu$m (section \ref{mixtures}). 

The extremely high degree of agreement between the Garnett EMT and the heterogeneous mixture shown in figure \ref{fig:mix_garn} was initially a surprise to us, given the presence of materials of high refractive index (see Appendix C). However, we believe this agreement actually validates not only the Garnett EMT itself (which hardly needs more validation) but also our numerical implementation of it. Consider the fact that for tiny particles with $r \ll \lambda$, as described in section \ref{effs}, $Q_e$ becomes proportional to the total volume or mass of material, regardless of the specific grain size distribution. This suggests that $Q_e$ also becomes independent of the configuration of the grains; they can be dispersed or combined into clumps, as long as the clumps themselves remain much smaller than the wavelength, without affecting the result. Indeed this is the configuration implicit in the Garnett-averaged results {\it for the Pollack size distribution}; tiny particles are rearranged into aggregates of still-tiny particles. The good agreement testifies to the validity of the Garnett EMT in calculating a single set of effective refractive indices (at each wavelength) from which to calculate $Q_e$. That this is not trivial is demonstrated by the poor agreement obtained from the intuitively simpler linear volume mixing approximation to the mean refractive index (Appendix C; blue curve in Figure \ref{fig:mix_garn}). Figure \ref{fig:MG_lin} in Appendix C shows in another way how badly volume mixing performs when even small amounts of high-refractive-index material are involved, as they are here. Of course, as particles grow larger, this equivalence will fail (a hint of divergence is seen at short wavelength in figure \ref{fig:mix_garn}). In this regime we will continue to rely on the Garnett EMT. 

\begin{figure*}[t]                                 
\centering                                                                   
\includegraphics[angle=0,width=3.2in]{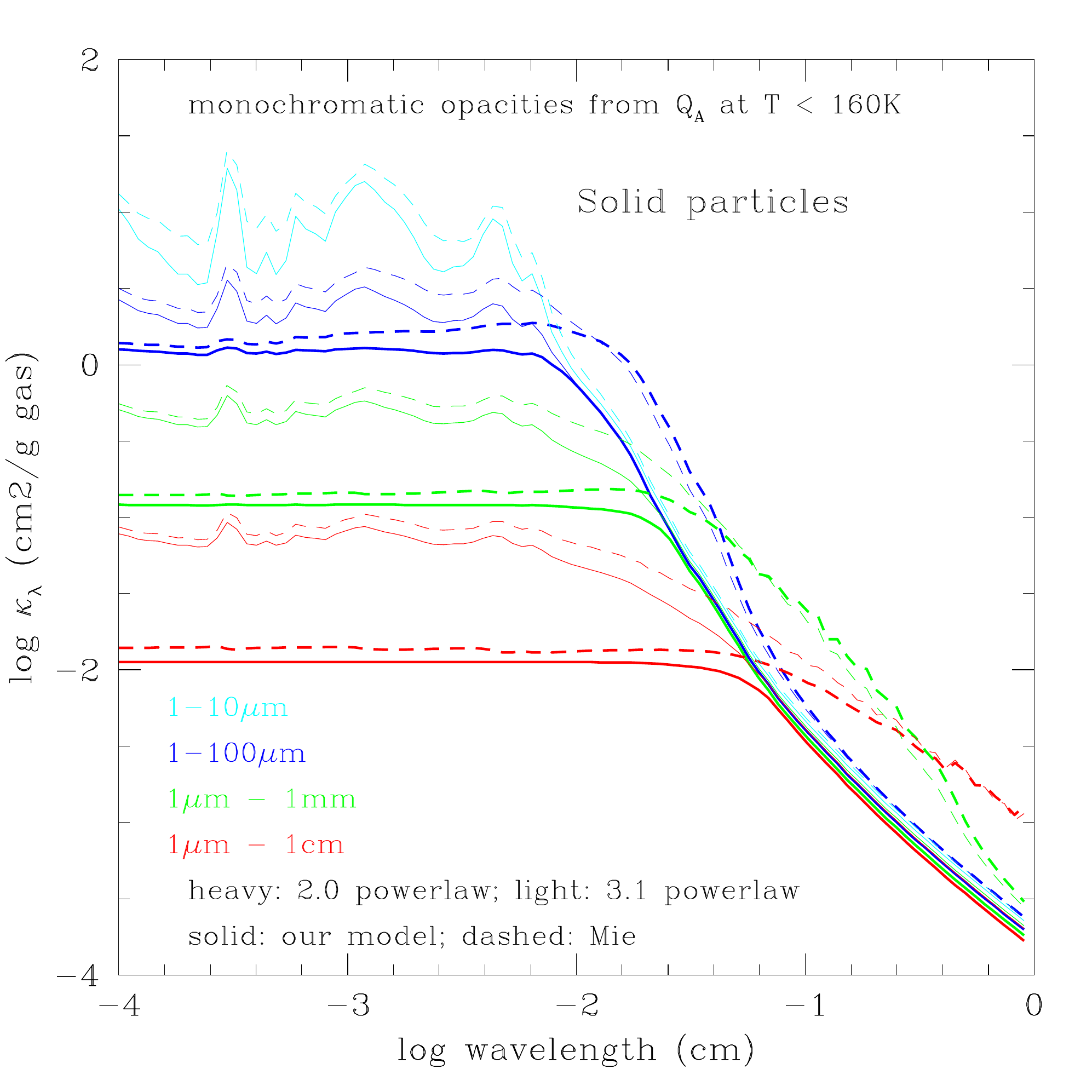}
\includegraphics[angle=0,width=3.2in]{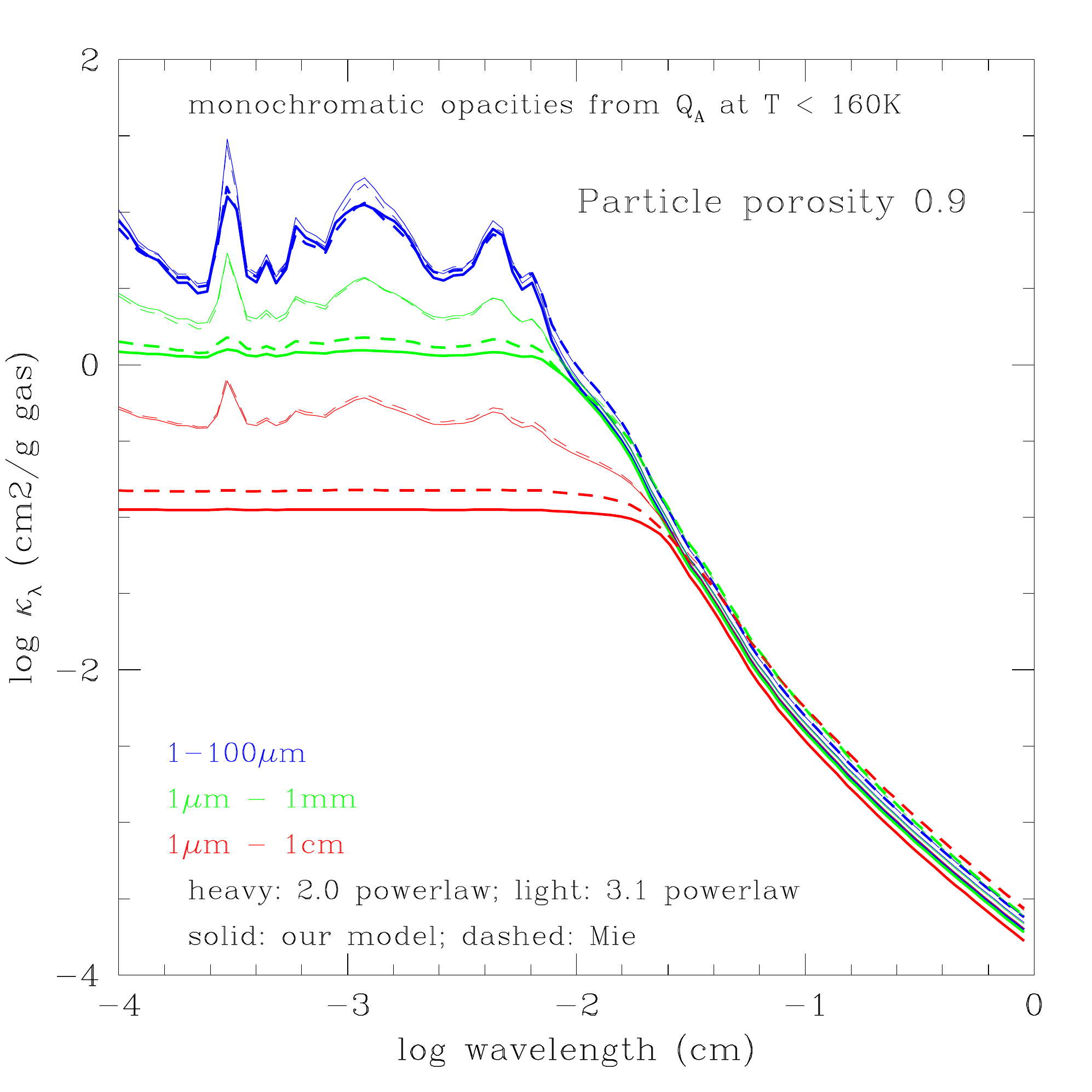}
\vspace{-0in}
\caption{Monochromatic opacity in pure absorption ({\it eg.} using only $Q_a$), such as would be relevant to interpreting mm-cm wavelength observations of disk fluxes. Two powerlaw distributions are shown, all of the form $n(r)=n_o r^{-s}$, where $n(r)$ is a particle number density per unit particle radius, with three different values of $r_{max}$. The particles are aggregates of mixed cosmic composition at 100K. Left: solid particles (porosity =0); right: porous particles (porosity =0.9).} 
\label{fig:SEDs}
\end{figure*}

\subsection{Monochromatic opacities and disk masses}\label{abs}
Perhaps the hardest challenge for our model is calculating monochromatic opacities for particle size distributions dominated by sizes comparable to the wavelength. Applications to wavelengths which are either much longer than the particle sizes in question ({\it eg.} most of figures \ref{fig:6panela} or \ref{fig:mix_garn}), or much shorter (geometrical optics), are very reliable and straightforward. To better determine the limitations of our model, we explore monochromatic opacity for broad size distributions where many of the particles are comparable to the wavelengths of greatest interest to mm-cm observations of disks. The results (figure \ref{fig:SEDs}) are comparable to those shown by Miyake and Nakagawa (1993; MN93) and D'Alessio et al (2001; D01). Because MN93 and D01 each adopt somewhat different choices for refractive indices and compositional regimes, we do not compare our results directly to theirs, but instead conduct our own Mie calculations using the Pollack et al (1994) refractive indices. D01 assumed a heterogeneous particle distribution (for instance, particles of pure ice, pure silicate, or pure troilite occur up to the mm-cm maximum sizes), whereas MN93 more typically assumed heterogeneous mixtures of aggregate particles with variable porosity, which we feel is more plausible under realistic conditions and also assume below. 

In figure \ref{fig:SEDs} we compare monochromatic profiles of the true absorption opacity $\kappa_{a,\lambda}$ from our model (equation \ref{eq:abscoef}) with full Mie calculations. MN93 have argued that scattering (which is important for {\it extinction} and thermal equilibrium modeling) is negligible in observations of this type and that pure absorption dominates thermal emission, and D01 agree. 

Results are shown for differential size distributions $n(r) = n_or^{-s}$, with $s$=2.0 and 3.1, with smallest radius of 1$\mu$m and largest radius $r_{max}$. The value $s=2$ (heavy curves) allows the larger particles to dominate the area and mass, while $s=3.1$ (light curves) gives a more equitable distribution of area across the range of sizes.  For each powerlaw slope, $r_{max}$ is varied from 10$\mu$m to 1cm (the Mie calculations bog down for larger sizes). In the left panel, we show results for solid particles (internal density = 1.38 g/cm$^3$). In the right panel, we show particles with 90\% porosity. The solid curves are from our model and the dashed curves are full Mie calculations. 

The agreement for porous particles is extremely good for the full range of sizes and size distributions. For solid particles (figure \ref{fig:SEDs} left), the Mie calculations exhibit an enhanced $Q_a$ in the resonance region, covering perhaps a decade of wavelength around $2 \pi r/\lambda \sim 1$ (see eg HT). This effect is difficult for a model lacking complete physical optics to reproduce. Such ``bumps" in $\kappa$ are visible in figure 5 of MN93, for the solid particle case, but lacking in their figure 8 (for 90\% porous particles), in good agreement with our results where all combinations of $r_{max}$ and powerlaw slope reach the same long-wavelength asymptote rather quickly. The same ``bump" is seen, for instance, in figure 3 of Ricci et al (2010), where the particles are actually not very porous ($\sim$ 40\%?). It is this contribution which carries through at shorter wavelengths, leaving our opacities 20\% low or so relative to Mie values; this small difference, in a wavelength-independent regime, is probably insignificant for most purposes when the maximum particle size {\it is} probably close to the specific observing wavelength {\it and} the porosity is not large, as discussed more below. 

The more shallow 2.0 powerlaw, dominated in area and mass by particles at or near the upper size cutoff, abruptly transitions to wavelength-independent opacity at the wavelength where, roughly, $Q_a(r_{max},\lambda)=1$. At longer wavelengths, a universal curve is followed, characterized by the total mass in the system. Notice that, for the 2.0 powerlaws, the inflection point moves to shorter wavelengths and the short-wavelength opacity increases as the upper radius cutoff decreases {\it or} as the porosity increases. This is a direct implication of equations \ref{eq:longqa} and \ref{eq:dielconst}, which together imply that $Q_a \propto r n_i$, and in most cases, in spite of the imperfection of the linear volume mixing model, $n_i$ is roughly proportional to $(1-\phi)$ (figure \ref{fig:MG_lin}). Meanwhile, for wavelengths shorter than the $Q_a(r_{max},\lambda)=1$ turnover, each particle or radius $r$ is less massive than a solid particle by the factor $(1-\phi)$, so the {\it number} of particles is larger (for a given total mass) by a factor $1/(1-\phi)$ (see also section \ref{por_eff} below). That is, fluffy cm-size particles have ten times the opacity of solid particles of the same radius (in the short-wavelength limit) because their 10 times lower mass per particle allows there to be ten times as many of them than the solid particles of the same size. Their mass per particle is still 100 times larger than that of a mm-size {\it solid} particle, but their cross-sectional area per particle is 100 times larger, so  the curve for porous, cm-radius particles lies on top of the curve for solid, mm-radius particles even at short wavelengths. This behavior can be seen tracing the differences between the heavy red and black curves, through solid, dashed, and dotted manifestations. 

The lightweight curves, for the steeper 3.1 powerlaw size distribution having the same radius limits, contain more small particles and thus show more spectral signatures at mid-infrared wavelengths. They also show broader slope transitions in $\kappa^a_{\lambda}$ around $Q_a(r_{max},\lambda)=1$.  Note the envelope of slope for {\it all} combinations of powerlaw, upper size limit, and porosity, for which $Q_a(r_{max},\lambda)=1$ has not been reached. Opacities (fluxes) at these wavelengths are independent of particle size, and thus capture all the mass even if particles have grown to cm size. Different powerlaw slopes or other details of the size distributions could lead to slightly different functional forms in the transition region, and can thus be constrained by high-quality observations. These comparisons of our model monochromatic opacities with full Mie theory provide further support for the Rosseland mean opacities which are based on them, certainly for porous particles, and illustrate the extent of the limitations for solid particles. In applications such as mm-cm monochromatic SED slope analysis, with the goal of determining largest particle sizes, full Mie theory should be used, and are not burdensome here because the wavelengths of interest are not much smaller than the particle sizes of interest. 
\begin{figure*}[t]                                 
\centering                                                                   
\includegraphics[angle=0,width=0.65\textwidth]{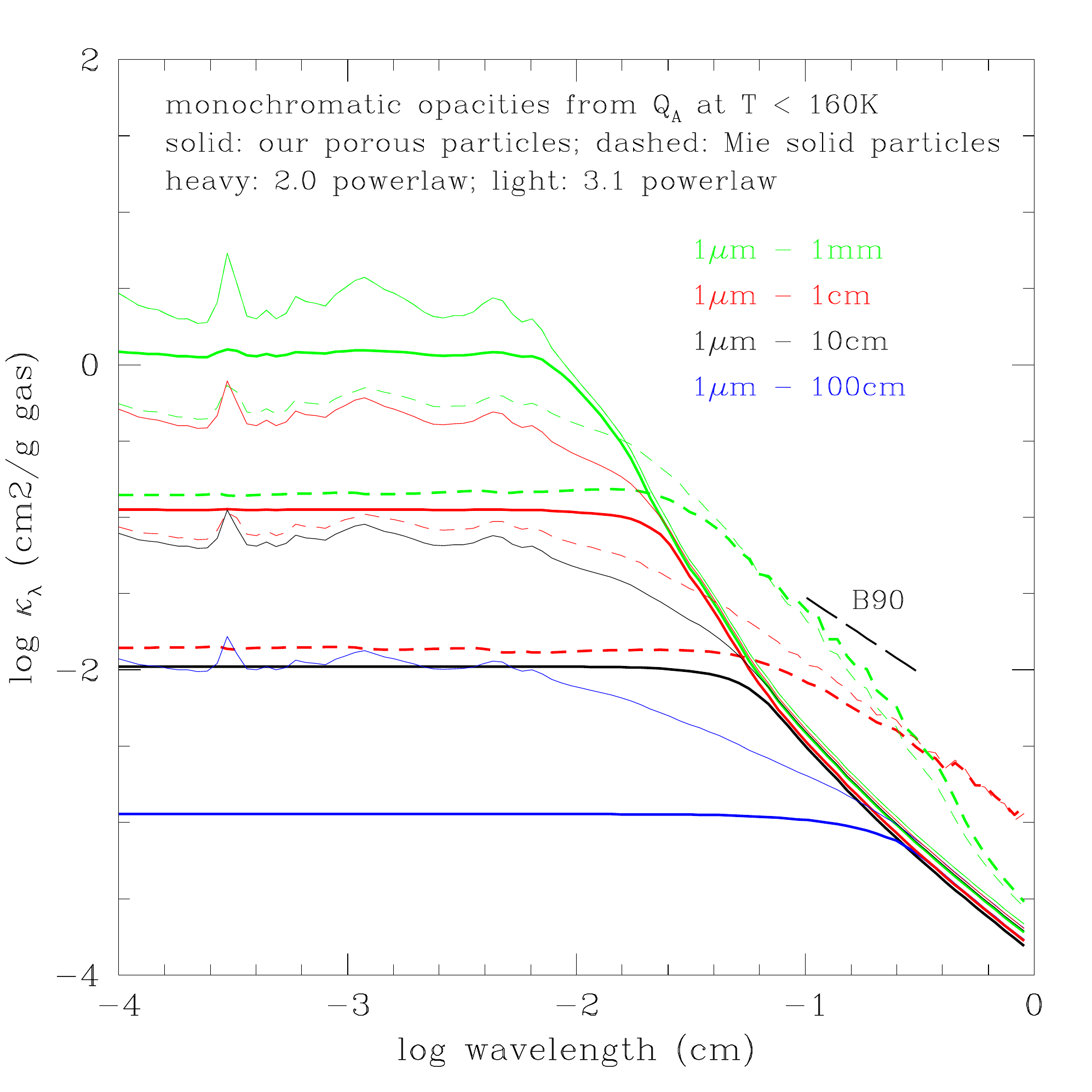}
\vspace{-0in}
\caption{Monochromatic opacity, comparing Mie calculations for solid particles (dashed lines) with our model for porous particles (solid lines). Two powerlaw distributions are shown: $s$=2.0 (heavy lines) and $s$=3.1 (light lines), and the upper size is varied from 1mm to 100cm (the 10 and 100cm size results are not shown for the Mie case because the mm-wavelength slopes would be too flat and the calculations are onerous). The widely used opacity law of Beckwith et al (1990) is the dashed line labeled B90. In principle, meter-size, very porous aggregates could be compatible with the {\it slope} of the mm-wavelength opacities; however as discussed in the text we believe that solid, or nearly solid, cm-size particles are more realistic. } 
\label{fig:SED2}
\end{figure*}

These scalings with size and porosity are further illustrated in figure \ref{fig:SED2}, comparing our runs for porous particles, which agree well with Mie calculations but are easier to extend to large sizes, with actual Mie calculations for solid particles.  In figure \ref{fig:SED2} we show a widely used opacity rule (Beckwith et al 1990, Williams and Cieza 2011): $\kappa_{\nu} = 0.1(\nu/10^{12} {\rm Hz})^{\beta}$, with $\beta$=1. Note that the canonical Beckwith et al (1990) opacity {\it could} be fit by a suite of porous particles extending to 100 cm radius, especially with a slope of 3.1 (or perhaps slightly steeper). As noted by previous authors, cm-size solid particles also provide a fairly good match to the slope. All models that match the slope have opacity quantitatively smaller than the canonical B90 opacity, however, by a factor of at least several (also found by Birnstiel et al 2010) possibly suggesting larger inferred disk masses. 

In outer disks where gas densities are very low, particles of even mm size and low-to-moderate porosity are dynamic radial migrators under the influence of gas drag (Takeuchi and Lin 2005, Brauer et al 2008, Hughes and Armitage 2010, 2012, Birnstiel et al 2010, 2012) and observations apparently show radial segregation of gas and particles in more than one outer disk (Andrews et al 2012, P{\'e}rez et al 2012). Moreover, such particles also couple to the large, high-velocity eddies and would be expected to have fairly high collision velocities. In this regime, large particles, even with 90\% porosity, would collide at significant velocities, inconsistent with retaining their postulated high porosities. This is because the aerodynamic coupling of a particle to turbulence is determined by the product of its radius and density (see, eg, V{\"o}lk et al l980, Cuzzi and Hogan 2003, Dominik et al 2007, Ormel and Cuzzi 2007). Thus it seems more plausible that the particles matching the B90 opacity in shape (or something like it) {\it are indeed} moderately compact, cm-radius, objects, and not meter-size, high-porosity puffballs; thus full Mie theory is needed for analyses of these observations.

\subsection{Rosseland mean opacity: effects of size and porosity}\label{por_eff}

\begin{figure*}[t]                                 
\centering                                                                   
\includegraphics[angle=0,width=3.2in,totalheight=2.5in]{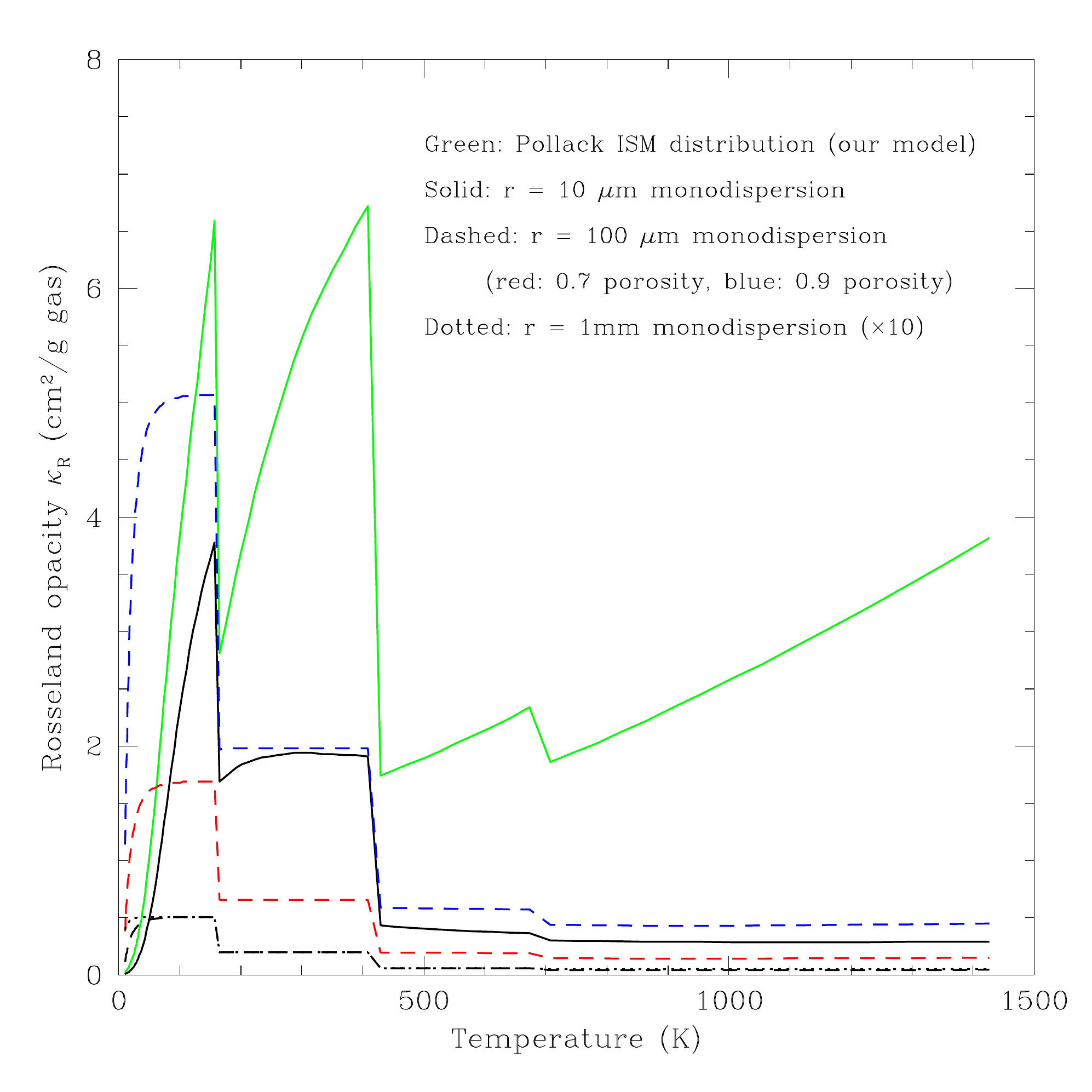}
\includegraphics[angle=0,width=3.2in,totalheight=2.5in]{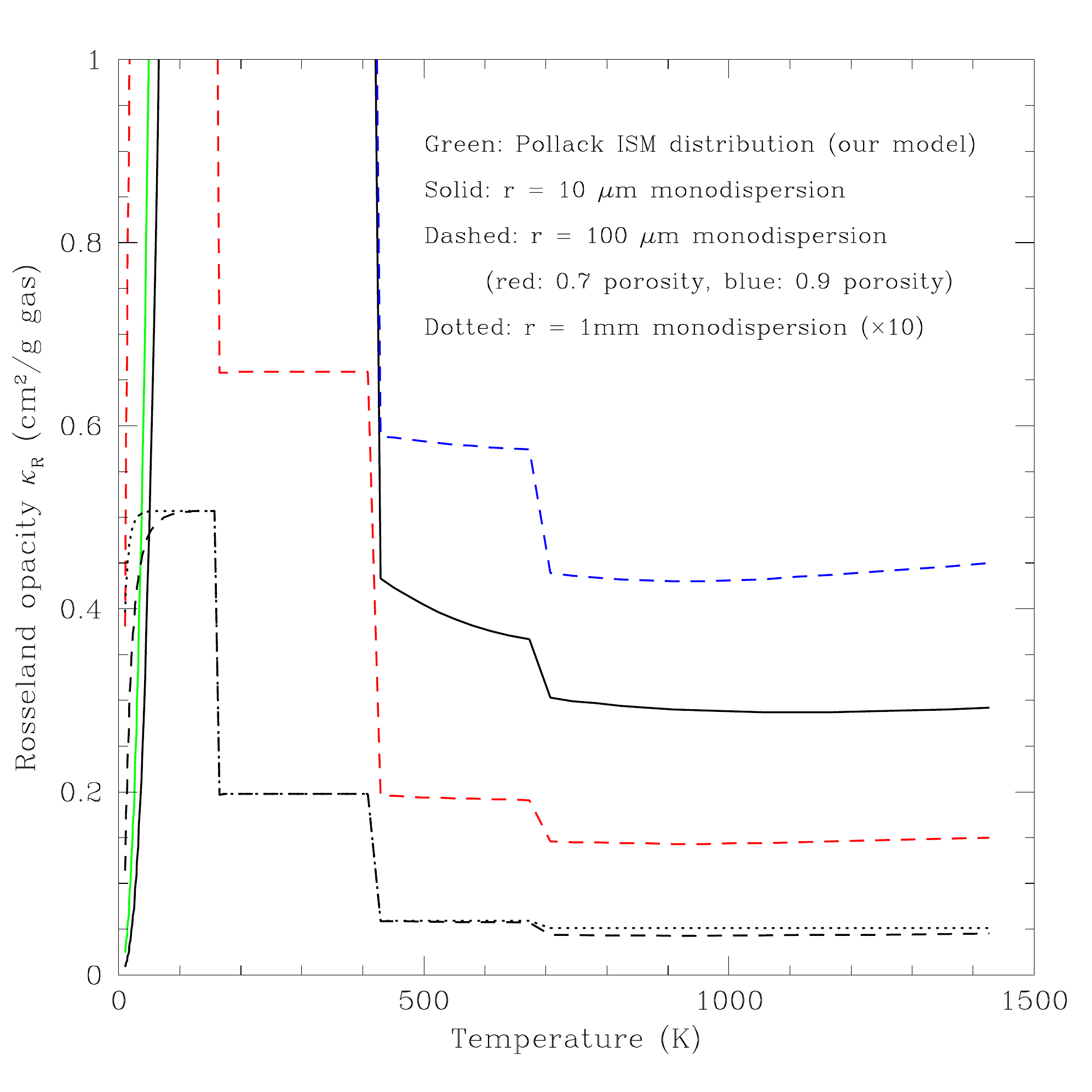}
\vspace{-0in}
\caption{Rosseland mean opacities for various particle sizes and porosities, all calculated using our  approach. Green: heterogeneous ISM distribution of Pollack et al (1994); black curves: solid aggregate particles of no porosity with the same mass; solid: 10 $\mu$m radius monodispersion; Dashed: 100 $\mu$m radius monodispersion; dotted: 1000 $\mu$m radius monodispersion (multiplied by 10). Also shown are red: 100 $\mu$m radius monodispersion with the same mass but porosity of 70\%; blue: 100 $\mu$m radius monodispersion with the same mass but porosity of 90\% (discussed in section 5.2).} 
\label{fig:rosscomp}
\end{figure*}

\begin{figure*}[t]                                 
\centering                                                                   
\includegraphics[angle=0,width=3.2in,totalheight=2.5in]{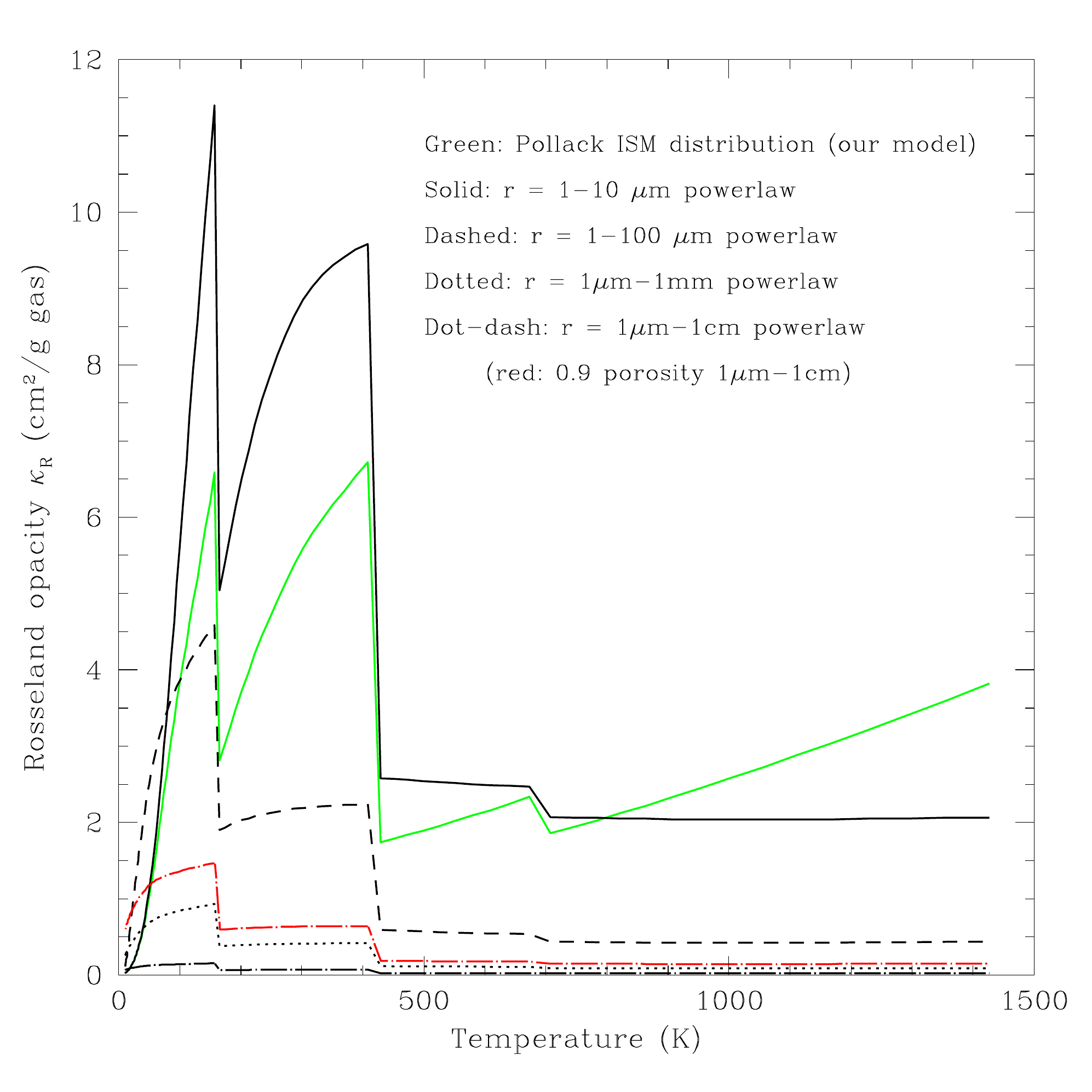}
\includegraphics[angle=0,width=3.2in,totalheight=2.5in]{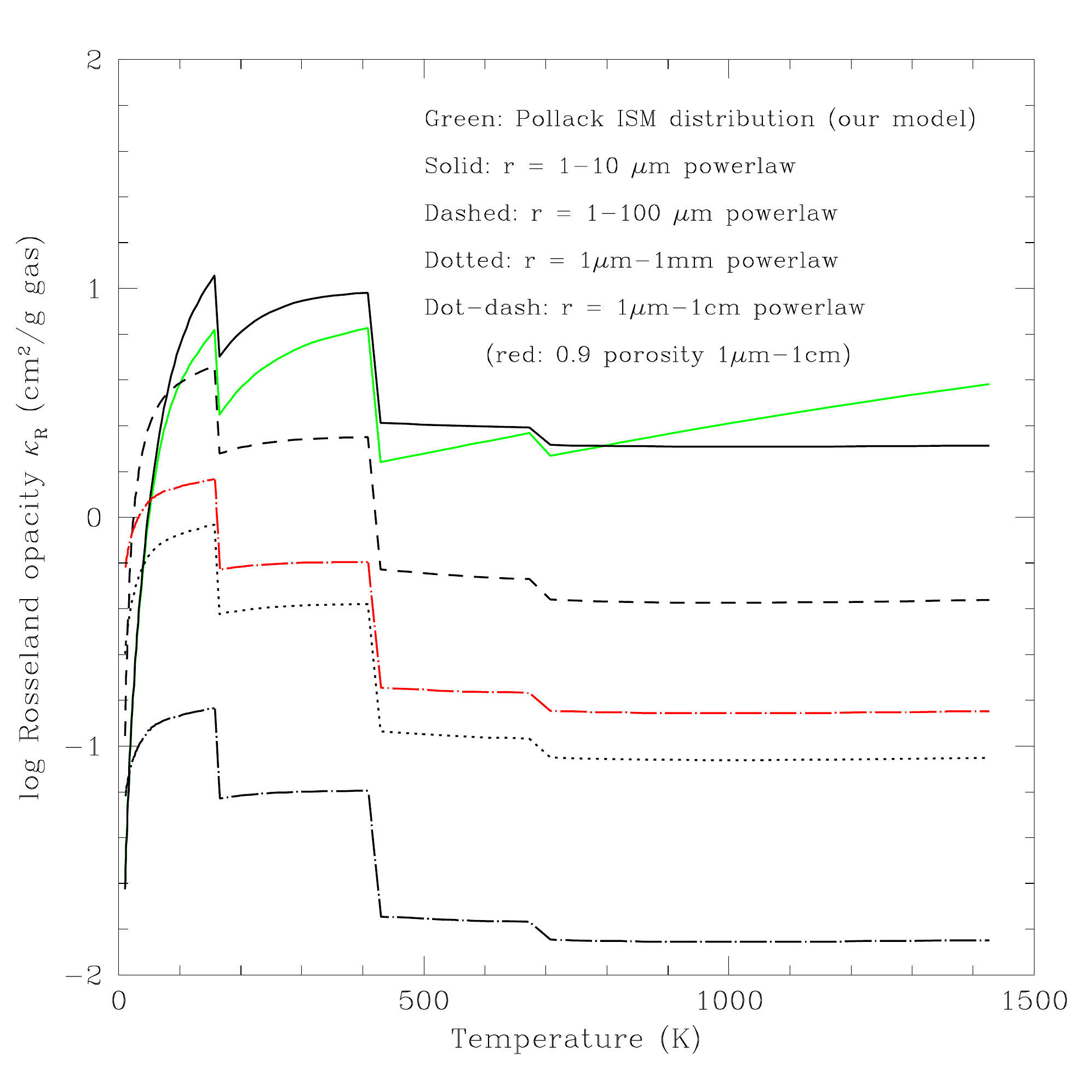}
\vspace{-0.1in}
\caption{Rosseland mean opacities for the same powerlaw size and porosity distributions as used in figure \ref{fig:SEDs}. Both linear and logarithmic forms are presented for the same results. Green: heterogeneous ISM distribution of Pollack et al (1994); black curves: solid aggregate particles of no porosity with the same mass and powerlaw size distributions of differential slope 3.1; solid: 1-10 $\mu$m radius; Dashed: 1-100 $\mu$m radius; dotted: 1$\mu$m -1mm radius; dot-dash: 1$\mu$m -1cm radius (shown are red: 1$\mu$m -1cm radius with porosity of 90\%. } 
\label{fig:ross_pl}
\end{figure*}

In this section we continue to assume well-mixed aggregates of material in each particle and extend our calculations of Rosseland mean opacities to particles with a wider range of size and porosity. First we show how growth affects the Rosseland mean opacities introduced in section 3.1 and the Appendix. Clearly, for particle size much larger than the wavelength, the opacity varies as the ratio of cross section to mass, or as $1/r$ (see Appendix of Pollack et al 1994 for a longer exposition). As shown by Pollack et al (1985) and Miyake and Nakagawa (1993), growth from microns to centimeters implies a decrease in $\kappa_R$ by four orders of magnitude, dwarfing any uncertainties regarding the actual composition of the particles and rendering the small differences seen in figures \ref{fig:6panela} - \ref{fig:mix_garn} somewhat moot. We illustrate the effect using figures \ref{fig:rosscomp} and \ref{fig:ross_pl}. The size distributions in figure \ref{fig:rosscomp} are monodispersions, which we have cautioned about previously, but in these calculations the large range of wavelengths over which efficiencies are integrated mimics the effects of a broad size distribution. 

Particle growth from the ISM distribution of Pollack et al (containing many submicron grains which are effective at blocking shortwave radiation characteristic of the higher temperatures shown), rapidly decreases the Rosseland opacity, even when growth is only to radius of 10$\mu$m (solid black curve). The decrease continues as particles grow to 100$\mu$m (dashed black) and 1mm (dotted black, multiplied by 10) radii. The opacity curves for the larger particles have a characteristic stepped appearance, with the steps representing changes in solid mass fraction at different evaporation temperatures. The opacity is nearly constant within a step, because particles much larger than a wavelength are in the constant-$Q_e$ regime independent of wavelength. Comparing the black dashed and dotted curves, where the dotted curve is 10 times the value of $\kappa_R$ for 1 mm radius particles, shows that the $1/r$ scaling is nearly exact for these sizes. The scaling is only slightly less easily explained between the 10$\mu$ m and 100$\mu$m radius solid particles. At the lower temperatures (longer wavelengths) the 10$\mu$m particles have not yet reached the geometrical optics limit, which occurs roughly for temperatures higher than 300K where the blackbody peak wavelength is around 10$\mu$m. Thus, the scaling between these two sizes is less than a factor of ten by an amount that depends on temperature (wavelength).

Figure \ref{fig:rosscomp} also shows the expected effects of porosity (section \ref{abs}); the red and green curves, for 100$\mu$m radius monodispersions of particles having porosities of 70\% and 90\% respectively, show good agreement with the expected scaling by $1/(1-\phi)$; that is, to conserve mass if the particle density decreases to $(1-\phi)\overline{\rho}$, we must increase their {\it number density} by a factor of $1/(1-\phi)$. In this regime where the particles are already much larger than the most heavily weighted wavelengths in question (except at the far left of the plot), the $Q_e$ per particle does not change, so the net opacity of the ensemble increases by a factor of $1/(1-\phi)$. So, the particle growth and porosity effects are robust, easily explained using simple physical arguments, and captured by the model. 

The Rosseland mean opacities for several {\it wide powerlaw} size distributions are shown in figure \ref{fig:ross_pl}. The behavior is similar overall to behavior seen previously and explained above. These powerlaws have a smallest particle radius (1 $\mu$m) which is slightly larger than the typical size on the Pollack et al size distribution, so $\kappa_R$ is somewhat larger at low temperatures (long effective wavelengths) but slightly lower and flatter at high temperature (short effective wavelengths). Powerlaws extending to larger size reduce the overall opacity at higher temperature, at least, by tying up more mass in larger particles. Porosity increases the overall opacity everywhere for the powerlaws extending to the larger sizes, because the dominant wavelengths involved are generally tens of microns or less. However, some wrinkles in the details, and differences between figures \ref{fig:rosscomp} and \ref{fig:ross_pl}, arise from the scattering terms due to $Q_s(1-g)$ in $\kappa_R$, which become important for particles with low $n_i$ when the particle size and wavelength are comparable. 

It is these drastic decreases in opacity with grain growth (Movshovitz and Podolak 2008) that led Movshovitz et al (2010) to conclude that gas giant formation could have happened much earlier than previous estimates which had {\it already} assumed an arbitrary 50x cut in opacity relative to the Pollack et al (1994) baseline (Hubickyj et al 2005, Lissauer et al 2009). The reason is that growth in the most important part of the radiative zone proceeds to {\it cm-size}, implying opacities even 10x smaller there than the {\it smallest} (for solid mm-size particles) shown in figures \ref{fig:rosscomp} and \ref{fig:ross_pl}. If the particles are porous, their opacity could be increased, however, as shown in those figures. This would seem to be an example of how the destiny of the great can be determined by the behavior of the small. 

\section{Conclusions} We outline a very simple, closed-form radiative transfer model which incorporates and connects well-understood asymptotic behavior for particles smaller than and larger than the wavelength. The approach is simple enough to provide good physical insight and to include in evolutionary models requiring radiative transfer. The model is easily adapted to arbitrary combinations of particle size, composition, and porosity across the range of plausible protoplanetary nebula and exoplanet cloud particle properties (excepting highly elongated particles), and yields values of Rosseland mean opacity which are in good agreement with more sophisticated but more time consuming Mie or DDA calculations. Planck opacities are even simpler to calculate (they are straight Blackbody-weighted means over wavelength). We illustrate the significant roles of particle growth and porosity in determining opacity. The model is {\it not} recommended even for Rosseland opacities in cases where ensembles of large, pure metal particles are expected. The method even gives very good approximations to {\it monochromatic} opacities {\it unless} the particles are solid and the specific wavelength of interest is comparable to the dominant particle size, where the model is unable to track ``resonance" behavior of the absorption efficiency. Thus for detailed spectral index analysis of mm-cm wavelength protoplanetary nebula emission spectra, in cases where particles in the mm-cm radius range might be solid and contribute significant mass and area, full Mie theory should be used. It appears that canonical mm-wavelength spectral slopes are more plausibly explained by solid, cm-size particles than larger, but more porous, aggregates. 

To summarize, the model consists of equations \ref{eq:longqa} and (\ref{eq:longqs} or  \ref{eq:raygans}), along with equation \ref{eq:qcorr} as constrained by 
equations \ref{eq:g3}-\ref{eq:trunc}. These equations lead to a 
monochromatic opacity $\kappa_{e,\lambda}$, which is then used to calculate a Rosseland mean $\kappa_R$ using equation \ref{eq:rossmean}. For calculation of mm-cm SEDs, only $\kappa^a_R$ should be used (equation \ref{eq:abscoef}). Our treatment can assume either a heterogeneous ``salt and pepper" particle size and compositional mix, as in Pollack et al 1994 or D'Alessio et al 2001 (with equations \ref{eq:kap_ej}-\ref{eq:heterosum}), or a distribution of internally-mixed, homogeneous aggregate particles having arbitrary porosity (using equations \ref{eq:eprime} - \ref{eq:nvsepsilon} in equations \ref{eq:longqa}-\ref{eq:trunc}, and integrating with equation \ref{eq:homog}), as for instance modeled by Miyake and Nakagawa (1993). 

\vspace{0.2 in}

{\bf Acknowledgements:} JNC benefitted greatly from many discussions of radiative transfer with Jim Pollack over the years. We thank Ted Roush for providing tabular values of the refractive indices used by Pollack et al (1994). We thank Pat Cassen,  Ke Chang, Tom Greene, Lee Hartmann, Stu Weidenschilling, and Diane Wooden for helpful conversations. We thank Kees Dullemond and Naor Movshovitz for encouragement to make this work more widely available. We thank our reviewer for helpful comments. JNC and PRE were supported under a grant from the Origins of Solar Systems Program. SSD was partially supported by Grants from the LASER and Astrobiology programs.

\section*{References}
 
Ackerman A. S. and Marley M. S. (2001) ApJ, 556, 872

Amit, L.; and Podolak, M. (2009) Icarus, 203, 610 

Andrews, S. M.; Williams, J. P. (2007) ApJ 659, 705 

Andrews, S. M.; Wilner, D.J.; Hughes, A. M. et al (2012) ApJ, 744, article id. 162
	
Beckwith, S. V. W. et al. (1990) AJ 99, 924 

Beckwith S. V. W., Henning T., and Nakagawa Y. (2000) In ``Protostars and Planets IV" (V. Mannings et al., eds.), p. 533. Univ. of Arizona, Tucson.

Birnstiel, T.; Dullemond, C. P.; Brauer, F. (2010) A\&A 513, id.A79

Birnstiel, T.; Klahr, H.; Ercolano, B. (2012) A\&A 539, id.A148 

Blum J. (2010) A\&A 10, 1199 
	
Bodenheimer, P.; Hubickyj, O.; Lissauer, J. J. (2000) Icarus, 143, 2

Bohren C. and Huffman D. R. (1983) Absorption and Scattering of Light by Small Particles. Wiley, New York.

Brauer, F.; Dullemond, C. P.; Henning, Th. (2008) A\&A 480, 859 

Currie, T.; Burrows, A.; Itoh, Y., et al (2011) ApJ 729, article id. 128
	
Cuzzi, J. N.; Hogan, R. C. (2003) Icarus 164, 127 

Cuzzi, J. N., Pollack, J. B. (1978) Icarus, 33, 233 
			
Cuzzi, J. N.; Weidenschilling, S. J. (2006) in ``Meteorites and the Early Solar System II", D. S. Lauretta and H. Y. McSween Jr. (eds.), University of Arizona Press, Tucson, p.353 

D'Alessio, P.; Calvet, N.; Hartmann, L. (2001) ApJ 553, 321 
	
D'Alessio, P., N. Calvet, L. Hartmann, et al (2006) ApJ 638, 314 

D'Alessio, P.; Calvet, N.; Hartmann, L.; et al (1999) ApJ 527, 893 

de Kok R. J., Helling C., Stam D. M., et al (2011) A\&A 531, A67.
	
Dominik C., Blum J., Cuzzi J. N., and Wurm G. (2007) In ``Protostars and Planets V" (B. Reipurth et al., eds.),  Univ. of Arizona, Tucson.

Dominik C. and Tielens A. G. G. M. (1997) ApJ 480, 647.

Donn, B. D. (1990); A\&A 235, 441 

Draine, B. T.; Goodman, J. (1993) ApJ 405, 685 

Draine B. T. and Lee H. M. (1984) ApJ, 285, 89 
	
Dullemond, C. P.; van Zadelhoff, G. J.; Natta, A. (2002);A\&A 389, 464-474

Fabian, D.; Henning, T.; J{\"a}ger, C. et al (2001)  A\&A 78, 228 
	
Ferguson, J. W.; Alexander, D. R.; Allard, F. et al (2005) ApJ 623, 585 

G{\"u}ttler, C.; Blum, J.; Zsom, A. et al (2010) A\&A 513, id.A56  

Hansen J. E. and Travis L. (1974) Sp. Sci. Rev., 16, 527 

Helled, R.; Bodenheimer, P. (2011) Icarus, 211, 939 

Helling, Ch.; Oevermann, M.; L{\"u}ttke, M. J. H. et al (2001) A\&A 376, 194 

Helling, Ch.; Ackerman, A.; Allard, F. et al (2008) MNRAS 391, 1854 

Henning, Th.; Il'In, V. B.; Krivova, N. A. et al (1999) 136, 405 
	
Henning, T.; Stognienko, R. (1996) A\&A 311, 291 

Hubickyj, O.; Bodenheimer, P.; Lissauer, J. J. (2005) Icarus, 179, 415 
	
Hughes, A. L. H.; Armitage, P.J. (2010) ApJ 719, 1633 

Hughes, A. L. H.; Armitage, P.J. (2012) MNRAS 423, 389 

Irvine W. M. (1975) Icarus, 25, 175 

Isella, A., J. M. Carpenter, and A. I. Sargent (2009) ApJ 701, 260 

Kaltenegger L., Traub W. A., and Jucks K. W. (2007) ApJ 658, 598 

Kitzmann D., Patzer A. B. C., von Paris P., et al. (2010) A\&A 511, A66.

Landau, L. D.; Lifshitz, E. M. (1960) Electrodynamics of continuous media; Oxford: Pergamon Press

Lindsay, S. S.; Wooden, D. H.; Harker, D.E. et al (2013) ApJ 766, article id. 54 

Lissauer, J. J.; Hubickyj, O.; D'Angelo, G.; Bodenheimer, P. (2009) Icarus, 199, 338 
	
Lodders K. and Fegley B. (2002) Icarus, 155, 393.

Madhusudhan N., Burrows A., and Currie T. (2011) ApJ 737, 34.

Marley M. S., Ackerman A. S., Cuzzi J. N., and Kitzmann D. (2013) In ``Comparative Climatology of Terrestrial Planets" (S. J. Mackwell et al., eds.), Univ. of Arizona, Tucson, in press; arXiv:1301.5627

Marley M., Gelino C., Stephens D., et al (1999) ApJ 513, 879 
	
Mathis, J. S.; Rumpl, W.; Nordsieck, K. H. (1977) ApJ 217,  425 

Meakin, P.; Donn, B. (1988) ApJ 329,  L39 

Min, M.; Hovenier, J. W.; Waters, L. B. F. M.; de Koter, A. (2008) A\&A 489, 135 
		
Miyake K. and Nakagawa Y. (1993) Icarus, 106, 20 

Mizuno, H., W. J. Markiewicz, and H. J. V\"{o}lk (1988) A\&A 195,
183 

Morley C., Fortney J., Marley M. et al (2012) ApJ 756, 172.

Morley C., Fortney J. J., Kempton E. M.-R. et al (2013) ApJ 775, article id. 33

Movshovitz, N., and M. Podolak, M. (2008) Icarus, 194, 368 

Movshovitz, N., Bodenheimer, P., Podolak, M., and Lissauer, J. J. (2010) Icarus, 209, 616 

Nakamoto, T. and Y. Nakagawa (1994) ApJ 421, 640

Natta, A.; Testi, L.; Calvet, N. et al (2007) in ``Protostars and Planets V", B. Reipurth, D. Jewitt, and K. Keil (eds.), University of Arizona Press, Tucson, 767 
	
Ormel, C. W.; Cuzzi, J. N. (2007) A\&A 466, 413 

Ormel, C. W.; Cuzzi, J. N.; Tielens, A. G. G. M. (2008) ApJ 679, 1588 
	
Ormel, C.; Okuzumi, S. (2013) ApJ 771, article id. 44
		
Ossenkopf V. (1991) A\&A 251, 210 

P{\'e}rez, L.M.; Carpenter, J. M.; Chandler, C. J. et al (2012) ApJL 760, article id. L17
	
Perrin, J.-M.; Lamy, P. L. (1990 ApJ, 364, 146 
	
Podolak, M. (2003) Icarus, 165, 428 
	
Pollack, J. B. and J. N. Cuzzi (1980) J. Atmos. Sci. 37, 868 

Pollack J. B., Hollenbach D., Beckwith S., et al. (1994) ApJ 421, 615 

Pollack, J. B., C. P. McKay, and B. M. Christofferson (1985) Icarus, 64, 471 

Purcell, E.M.; Pennypacker, C. R. (1973) ApJ 186, 705 

Rannou, P.; McKay, C. P.; Botet, R.; Cabane, M. (1999) Plan. and Sp. Sci.,  47, 385 

Ricci, L.; Testi, L.; Natta, A.; Brooks, K. J. (2010) A\&A 521, id.A66
	
Schraepler, R.; Blum, J.; Seizinger, A.; Kley, W. (2012) ApJ 758, article id. 35

Semenov, D.; Henning, Th.; Helling, Ch. et al (2003) A\&A 410, 611 

Stognienko R., Henning Th., and Ossenkopf V. (1995) A\&A 296, 797 

Sudarsky, D.; Burrows, A.; Hubeny, I. (2003) ApJ 588, 1121 

Sudarsky, D.; Burrows, A.; Pinto, P. (2000) ApJ 538, 885 

Takeuchi, T.; Lin, D. N. C. (2005) ApJ 623, 482 
	
Tanner, D. B. (1984) Phys. Rev. B., 30, 1042 

Tsuji, T. (2002) ApJ 575, 264 

van de Hulst H. C. (1957) Light Scattering by Small Particles.
Wiley and Sons, New York. 

van de Hulst H. C. (1980) Multiple Light Scattering, Vols. 1 and
2. Academic Press, New York.

Vasquez M., Schreier F., Gimeno Garc{\'i}a S. et al (2013) A\&A 557, id.A46
	
V{\"o}lk, H. J.; Jones, F. C.; Morfill, G. E.; Roeser, S.; (1980)  A\&A 85, 316 
	
Voshchinnikov N. V., II'in V. B., and Henning Th. (2005) A\&A 429, 371 

Voshchinnikov N. V., II'in V. B., Henning Th., and Dubkova D. N. (2006) A\&A 445, 167 

Weidenschilling, S. J. (1988) in ``Meteorites and the early solar system" Tucson, AZ, University of Arizona Press, 348-371.J. Kerridge and M. Matthews, eds. 
	
Weidenschilling, S. J. (1997) Icarus, 127, 290 

Williams, J. P.; Cieza, L. A. (2011) ARAA, 49, 67 
	
Wright E. L. (1987) ApJ 320, 818 

Zsom A., Ormel C. W., Guettler C., Blum J., and Dullemond C. P. (2010)  A\&A 513, A57.

Zsom, A.; Kaltenegger, L.; Goldblatt, C. (2012) Icarus, 221, 603

\section*{Appendix A: Rosseland Mean Opacity}
While derivations of the Rosseland mean opacity can be found in various sources, we provide here a quick derivation of $\kappa_R$ because it is surprisingly difficult to find good ones, and we make reference to some of its specific aspects. Essentially one derives an expression for the monochromatic flux in a high-opacity thermal radiation field, then integrates over frequency, and from this identifies the associated mean opacity. Start with the radiative transfer equation for monochromatic intensity  $I_{\nu}$ in a plane layered medium: $\mu dI_{\nu}/d\tau = S_{\nu} - I_{\nu}$ where $S_{\nu}$ is the source function and $\tau_{\nu}$ is monochromatic optical depth measured normal to the layer, and $\mu= {\rm cos}\theta$ where $\theta$ here is the angle from the layer normal and $d\tau_{\nu} = \kappa_{\nu} dz$ where $dz$ is an increment of thickness in the layer.  In a medium of high optical depth where thermal radiation dominates everything else, $I_{\nu} \approx S_{\nu} \approx B_{\nu}$ where $B_{\nu}$ is the Planck function. Then we can rewrite the above equation to first order as $I_{\nu} = B_{\nu} - dB_{\nu}/d\tau$. We then determine the local energy {\it flux} through the layer $F_{\nu} = 2\pi \int_{-1}^1 I_{\nu} \mu d\mu$. In the presence of the gradient derived above, substituting for $I_{\nu}$:
\begin{equation}
F_{\nu} = 2 \pi \left[ \int_{-1}^1 B_{\nu}\mu d\mu -
                       \int_{-1}^1{ \mu dB_{\nu} \over\kappa_{\nu} dz} \mu d\mu  \right].
\end{equation}
The first integral vanishes because $B_{\nu}$=constant; the flux becomes
\begin{equation}
F_{\nu} = -\frac{2 \pi}{\kappa_{\nu}} \frac{dB_{\nu}}{dz}\int_{-1}^1\mu^2d\mu
        = -\frac{4 \pi}{3 \kappa_{\nu}}\frac{dB_{\nu}}{dz}. 
\end{equation}
The standard trick is to set $dB_{\nu}/dz = (dB_{\nu}/dT)(dT/dz)$. The monochromatic flux is then integrated over frequency, after rearranging terms:
\begin{equation}
F = \int F_{\nu} d\nu = -\frac{4 \pi}{3}\frac{dT}{dz}\int \frac{1}{\kappa_{\nu}}\frac{dB_{\nu}}{dT}d\nu.
\end{equation}
One then merely asserts that the frequency integrated flux can be written in the same form, except with a mean opacity $\kappa_R$:
\begin{equation}
F = -\frac{4 \pi}{3}\frac{dT}{dz}\frac{1}{\kappa_R}\int \frac{dB_{\nu}}{dT}d\nu;\end{equation}
and after setting the two expressions equal, we obtain the definition of $\kappa_R$:
\begin{equation}\label{eq:rossmean}
\frac{1}{\kappa_R} = \frac{\int \frac{1}{\kappa_{\nu}}\frac{dB_{\nu}}{dT}d\nu}{\int \frac{dB_{\nu}}{dT}d\nu};
\end{equation}
essentially, we are obtaining the weighted average of $1/\kappa_{\nu}$, thereby emphasizing spectral regions where energy ``leaks through", and where the integral of the weighting function $dB_{\nu}/dT$ in the denominator may, if we like, be further simplified as 
$d/dT(\int B_{\nu}d\nu) = 4\sigma_{SB}T^3$, where $\sigma_{SB}$ is the Stefan-Bolzmann constant. 

\section*{Appendix B: Metal Particles} 
The small particle expansions of DL84, leading to our primary equations \ref{eq:longqa} and \ref{eq:longqs}, incorporate only electric dipole terms. Similar expressions can be derived for magnetic dipole terms, which are more cumbersome in their full glory (see eg Tanner 1984 or Ossenkopf 1991) and are usually approximated. For instance, 
DL84 (equation 3.27), citing Landau and Lifshitz (1960 sections 45, 72, and 73), give a handy first-order correction factor for $Q_a$ with the caveat that it is valid for small $x$, but without specific guidance as to what is ``small".  The correction is simply $Q'_a(r,\lambda) = Q_a(r,\lambda)(1 + F)$, where
\begin{equation}\label{eq:dlfac}
F = \left( \frac{2 \pi r}{\lambda}\right)^2{(\epsilon_1+2)^2 +\epsilon_2^2 \over 90},
\end{equation}
and $\epsilon_1 = n_r^2 -n_i^2$ and $\epsilon_2=2 n_r n_i$ (equation \ref{eq:dielconst}). Similar expressions appear in Ossenkopf (1991) and Tanner (1984), and surely elsewhere. The factor $F$ gets very large when the refractive indices are large. In the case of iron metal, where $n_r$ and $n_i$ are nearly proportional to $\lambda$,  the spectral behavior of equation \ref{eq:longqa} is flattened from $x^{-3}$ nearly to $x^{-1}$, as can be seen by expanding terms in the large-refractive-index limit. 

This magnetic dipole correction is insignificant in the mixed-material, aggregate particle case, as it only enters when the refractive indices of a particle are $\gg \lambda/r$, so readers interested only in aggregate grains need not be concerned further with this term. Yet, for those who might be interested in clouds of metallic particles, the correction might be of interest. Unfortunately, it is formally only valid for particles that are small compared to the wavelength {\it inside} the particle, or equivalently the skin depth of the wave, which is given by $x \sqrt{\epsilon} \ll 1$, and for the large $\epsilon$ of metals (see figure \ref{fig:indices}) the allowed $x$ is much smaller than the usual condition $x=2 \pi r/\lambda \ll 1$. The correction is mentioned by Pollack et al (1994) but it is unclear from that paper just how it was used; their figure 2a shows results for metal particles having their standard size distribution in a wavelength range where it is not formally valid, and they give no comparison with Mie theory. We have found that if equation \ref{eq:dlfac} is blithely applied for $2\pi r/\lambda < 1$, well out of its formal domain of validity, and the ensuing $Q_a$ is subject to our overall constraint $Q_a < 1$), the agreement with Mie calculations for pure metal is improved from nonexistent to marginal (figure \ref{fig:6panela}). Tanner (1984) has found similar behavior, in that use of the ``approximation" outside of its formal domain of validity gives surprisingly better agreement with observed behavior than application of the more complete theory. This is not an argument for general acceptance of the approximation, and in any application where metal particles dominate the situation the complete Mie theory is probably required. Nevertheless, we have employed it in our model.

\section*{Appendix C: Garnett Effective Medium Theory}
In this appendix, we give an overview of the Garnett 
theory for the average complex dielectric constant ($\epsilon$) of an 
inhomogeneous medium. The reader is referred to Bohren and Huffman (1983) for a more complete exposition with background. 

A number of exhaustive and sophisticated studies have compared several different kinds of EMT models to rigorous, brute-force numerical Discrete Dipole Approximation (DDA) models (Perrin and Lamy 1990, Ossenkopf et al 1991, Stognienko et al 1995, Voshchinnikov et al 2005, 2006; see also Semenov et al 2003 and references therein); the differences are generally small and composition-dependent for small particles which can be treated using DDA (and much smaller than the huge differences due to growth which our model is primarily intended to capture). For instance, figure 16 of Voshchinnikov et al (2005) shows a nearly insignificant difference between Garnett and Bruggeman models relative to DDA calculations, in the regime where all scattering elements in the aggregates are truly small compared to the wavelength. Interestingly, they show that a model of their own device does a better job matching certain very porous aggregates containing monomers with a distribution of sizes (at least, at long wavelengths). We believe that both traditional EMT theories fail to match the ``size distribution of inclusions" DDA results of Voshchinnikov et al 2005 figure 16, because the aggregates contain numerous embedded wavelength-sized monomers, which violate the assumptions of {\it both} EMT models (some of the monomer inclusions have diameter as large as the wavelength). For this situation to occur at wavelengths, particles, and temperatures of interest for Rosseland mean opacities of our paper would require monomers in the few-to-tens of micron size, that are in turn embedded by assumption in much larger particles - which our model would treat in the geometric optics limit in any case. In sections 5.3 and the Conclusions, we note deviations from our model when a significant fraction of the particles of interest are wavelength-sized (we anticipate this to be a problem mostly for mm-cm wavelength observations). The behavior is similar to that seen in figure 16 of Voshchinnikov et al (2005). 

We derive the components of 
$\epsilon=\epsilon^{\prime}+i\epsilon^{\prime\prime}$ and convert the results 
to the perhaps more familiar complex refractive index $n^{\prime}+in^{\prime\prime}$. We 
then assess the realm of validity for a linear approximation of refractive 
index as a function of material density by comparing the two expressions for 
a  range of component materials.

\indent Interactions among different constituents of an inhomogeneous 
medium make the determination of an average dielectric constant problematic. 
The problem is generally insoluble, save by brute force (DDA) or approximation methods, 
and different approximations inevitably lead to different expressions.
As an example, one choice that can be found in the literature as 
far back as 1850 is the Rayleigh expression that relates the density and 
dielectric constant of a powder to that of the corresponding solid.
This result follows from the Clausius-Mosotti law that 
the quantity $\epsilon-1/\epsilon+2$ is proportional to the density of a 
material:\\
\begin{equation}\label{eq:rayleigh}
\frac{1}{\rho}\frac{\epsilon-1}{\epsilon+2} = 
\frac{1}{\rho_{o}}\frac{\epsilon_{o}-1}{\epsilon_{o}+2},
\end{equation}
where $\epsilon_{o},\rho_{o}$ are the complex dielectric constant and density 
of the solid material, 
and $\epsilon,\rho$ are those of the powder. This expression has been 
shown to be fairly accurate for powders made from various geological materials 
(e.g. Campbell and Ulrichs 1969). A more complete expression, which 
reduces to the Rayleigh formula in the case of one component, was derived by Maxwell 
Garnett (see, e.g. Bohren and Hoffman 1983). Garnett's model is that of  
inclusions of dielectric constant $\epsilon_{o}$ embedded 
in a homogeneous medium of dielectric constant $\epsilon_{m}$. In the simplest 
version, the inclusions are assumed to 
be identical in composition, but may vary in shape, size, and orientation. 
Assuming that all the inclusions are 
spherical, the expression for the average dielectric constant $\epsilon$ 
is given by
\begin{equation} \label{eq:diel}
\epsilon = \epsilon_m \left[ 1 + 3f 
 \left( \frac{\epsilon_o - \epsilon_m}{\epsilon_o + 2\epsilon_m} \right) 
 \left( 1 - f 
 \left( \frac{\epsilon_o - \epsilon_m}{\epsilon_o + 2\epsilon_m} \right)
 \right) ^{-1} \right],
\end{equation}

where $f$ is a mass fraction which is defined below. For $\epsilon_{m}=1$, 
this reduces to the Rayleigh formula with $f=\rho/\rho_{o}$ (for a more detailed
derivation, including nonsphericity effects, see Bohren and Hoffman 1983). 
The advantage of 
the  Garnett equation is that it can be generalized to a multiple 
component medium. The general result can be cast into the same form as (\ref{eq:diel}) 
except now $f$ and $\epsilon_{o}$ become $f_{j}$ and $\epsilon_{j}$, while the numerator 
and denominator of (\ref{eq:diel}) are summed over $j$ species (Bohren and Hoffman 1983, 
Sect. 8.5).

Noting that 
$\epsilon=\epsilon^{\prime}+i\epsilon^{\prime\prime}$, we expand (\ref{eq:diel}) 
into its complex parts, convert  to fractional form, and separate the real 
and imaginary components: 
\begin{equation}\label{eq:complex}
\epsilon=
\epsilon^{\prime}+i\epsilon^{\prime\prime} = \frac{1+ 2\sum_{j} f_{j}\sigma_{j} 
+ i6\sum_{j} f_{j}\gamma_{j}}{1 - \sum_{j} f_{j}\sigma_{j} - i3\sum_{j} 
f_{j}\gamma_{j}},
\end{equation}
where we have defined $\sigma_{j}$ and $\gamma_{j}$ to be
\begin{equation}\label{eq:sigj}
\sigma_{j}=\frac{(\epsilon^{\prime}_{j}-1)(\epsilon^{\prime}_{j}+2) + 
\epsilon^{\prime\prime 2}_{j}}{(\epsilon^{\prime}_{j}+2)^{2} + 
\epsilon^{\prime\prime 2}_{j}},
\end{equation}
and
\begin{equation}\label{eq:gamj}
\gamma_{j}=\frac{\epsilon^{\prime\prime}_{j}}{(\epsilon^{\prime}_{j}+2)^{2} + 
\epsilon^{\prime\prime 2}_{j}},
\end{equation}
respectively. We apply the complex conjugate to (\ref{eq:complex}),
determine the real and imaginary parts of $\epsilon$, and combine  like 
terms to get
\begin{equation}\label{eq:eprime}
\epsilon^{\prime}=\frac{1+\sum_{j} f_{j}\sigma_{j} - 2\sum_{i} 
\sum_{j} f_{i}f_{j}(\sigma_{i}\sigma_{j}+9\gamma_{i}\gamma_{j})}{1 - 2\sum_{j} 
f_{j}\sigma_{j}+\sum_{i} \sum_{j} f_{i}f_{j}(\sigma_{i}\sigma_{j}+ 
9\gamma_{i}\gamma_{j})}=\frac{N_{R}}{D},
\end{equation}
for the real part, and
\begin{equation}\label{eq:edoubleprime}
\epsilon^{\prime\prime}=\frac{9\sum_{j} f_{j}\gamma_{j}}{1 - 
2\sum_{j} f_{j}\sigma_{j}+\sum_{i} \sum_{j} f_{i}f_{j}(\sigma_{i}\sigma_{j}+ 
9\gamma_{i}\gamma_{j})}=\frac{N_{I}}{D}
\end{equation}
for the imaginary part. A simple check will show that this generalized  
Garnett formula reduces to equation (\ref{eq:rayleigh}), the Rayleigh formula, for $i=j=1$. \\

Conversion of equations (\ref{eq:eprime}) and (\ref{eq:edoubleprime}) to an expression in terms of refractive indices 
poses no great problem. Making the usual assumption that magnetic permeability is 
of order unity (see however Appendix C) we define
\begin{equation}
\epsilon^{\prime}+i\epsilon^{\prime\prime}=(n^{\prime}+in^{\prime\prime})^{2},
\end{equation}
where now, (\ref{eq:sigj}) and (\ref{eq:gamj}) become
\begin{equation}
\sigma_{j}=\frac{(n^{\prime 2}_{j}-n^{\prime\prime 2}_{j}-1)
(n^{\prime 2}_{j}-n^{\prime\prime 2}_{j}+2) + 
4n^{\prime 2}_{j}n^{\prime\prime 2}_{j}}{(n^{\prime 2}_{j} -
n^{\prime\prime 2}_{j}+2)^{2} + 4n^{\prime 2}_{j}n^{\prime\prime 2}_{j}},
\end{equation}
and
\begin{equation}
\gamma_{j}=\frac{2n^{\prime}_{j}n^{\prime\prime}_{j}}{(n^{\prime 2}_{j} -
n^{\prime\prime 2}_{j}+2)^{2} + 4n^{\prime 2}_{j}n^{\prime\prime 2}_{j}}
\end{equation}
respectively. Finally, we express the average refractive index 
of the inhomogeneous medium in terms of $\epsilon$ as
\begin{equation}\label{eq:nvsepsilon}
n^{\prime}+in^{\prime\prime}=\left[\frac{\sqrt{\epsilon^{\prime 2} + 
\epsilon^{\prime\prime 2}}+\epsilon^{\prime}}{2}\right]^{1/2} + 
i\left[\frac{\sqrt{\epsilon^{\prime 2} + 
\epsilon^{\prime\prime 2}}-\epsilon^{\prime}}{2}\right]^{1/2}
\end{equation}

The  Garnett formula defines the $f_{j}$ as the volume fraction 
of inclusions of species $j$ within a particle of some overall volume $V$ and total mass $M$; then the total solid volume fraction in the particle is $f = \sum_{j} f_j = 1-\phi$ where $\phi$ is the porosity of the particle.
We can determine $f_{j}$ in terms of mass fractions as follows: 
$v_{kj}$ is the $k$th volume element of species $j$ so that the volume 
fraction of inclusions of species $j$ is $f_{j}=\sum_{k}v_{kj}/V$. We define 
$\alpha_{j}$ as the mass of all component $j$ per unit nebula gas mass, and 
$\alpha=\sum_{j}\alpha_{j}$ is the mass fraction of all solids per unit nebula gas mass.
With these definitions in mind, 
\begin{equation}\label{eq:fj}
f_{j}=\frac{v_{j}}{V}=\frac{m_{j}}{\rho_{j}V}=
\frac{m_{j}\rho}{\rho_{j}M}=\frac{\rho}{\rho_{j}}\frac{\alpha_{j}}{\alpha}
 = \frac{(1-\phi) \overline{\rho} \alpha_j}{\rho_j \alpha}
\end{equation}
where $\rho$ is
the average mass density of a composite particle; $\rho$ can vary
from nearly zero to the ``solid" average value $\overline{\rho}$ at $\phi=0$. Using the last expression of equation (\ref{eq:fj}) in $\sum_j f_j = f = 1-\phi$ gives
\begin{equation}\label{eq:rhobar}
\overline{\rho}=\frac{\alpha}{\sum_{j} (\alpha_{j}/\rho_{j})}.
\end{equation}

Under certain conditions, a simple linear approximation for the refractive index at a particular 
wavelength $\lambda$ can greatly simplify calculations (see equations 22-23). Generally, such an 
approximation works well when $n$ is of order unity (such as visible 
wavelengths). 

Equations 22-23 follow from a simple, but approximate, extension of the so-called ``Wiener rule" (Voshchinnikov et al 2005): $\left<\epsilon\right> = \Sigma f_j \epsilon_j$, where $\epsilon_j$ are the dielectric constants of the different materials $j$ and the relation assumes lossless materials ($n_i=0)$. Eqns 22-23 are simply the analogous linear volume averages applied to ``moderate" refractive indices not far from unity, as is true for most silicates and ices.  Assuming that the Wiener rule can be applied to a case where $n_i \ne 0$, equations 22-23 can be derived in an approximate way, in the limit that $n_i \ll 1$ and $n_r = 1+\delta$ with $\delta^2 \ll 2\delta$. Given $\left<\epsilon\right> = \Sigma f_j \epsilon_j$, separate the real and imaginary parts of the sum recalling that $\epsilon  = (n_r + i n_i)^2$, giving $\left<\epsilon_i\right>= \Sigma f_j (2 n_i n_r) \sim 2\overline{n_r}\Sigma f_j n_{ji}$. The LHS of the equation can also be approximated as $2\overline{n_r}\left<n_i\right>$, so $\left<n_i\right> \sim \Sigma f_j n_{ji}$ as in equation (23). Similarly $\left<\epsilon_r\right> = \left<(1+\overline{\delta})^2\right> \sim 1+2\overline{\delta}$, and on the RHS $\Sigma f_j (1+\delta_j)^2 \sim \Sigma f_j (1+2\delta_j)$. Separating the sum on the RHS and noting that $\Sigma f_j =1$, we get $\left<\delta\right> \sim \Sigma f_j \delta_j$ as in equation (22). Of course, these {\it are} only approximations and valid only under the conditions stated. 

\begin{figure*}[t]                              
\centering                                                                   
\includegraphics[angle=0,width=3.2in,height=3.0in]{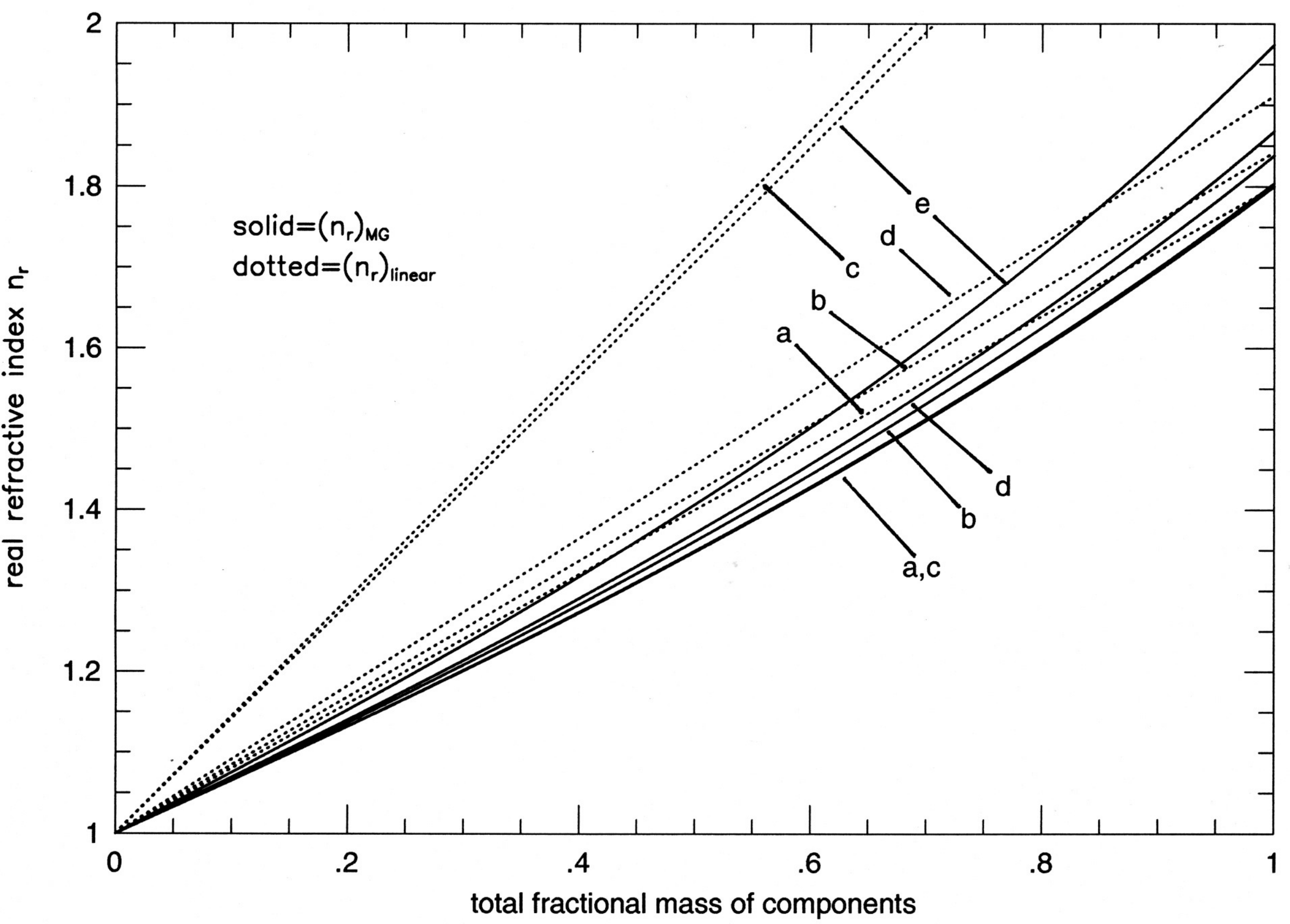}
\includegraphics[angle=0,width=3.2in,height=3.0in]{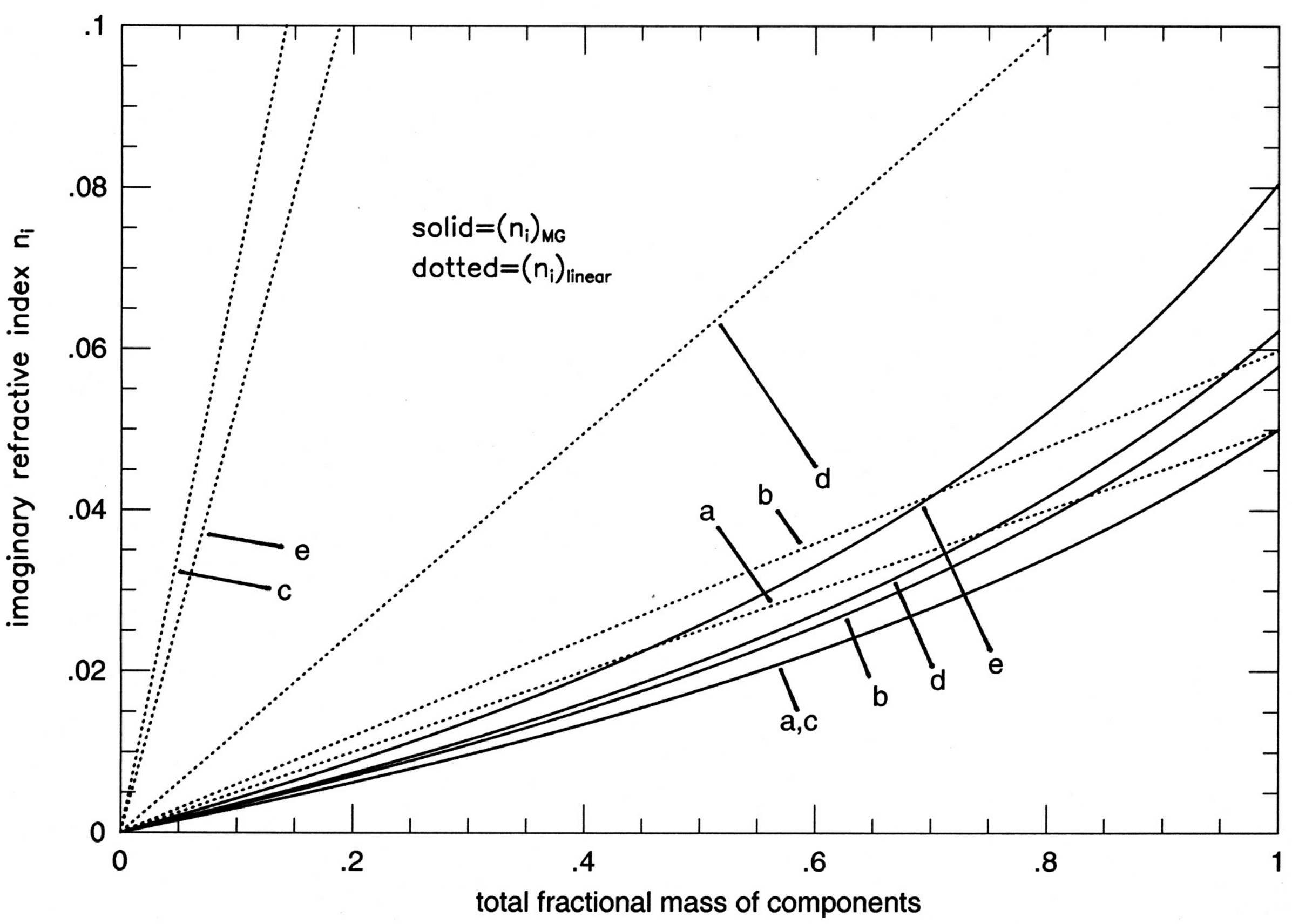}
\vspace{-0.0in}
\caption{Plots of the real (left)and imaginary (right) components of average 
refractive index of a material of density $\rho$ as a function of the density 
ratio $\rho/\overline{\rho}$.  Curves marked (a) 
correspond to pure ice. Curves (b,c) correspond to two component mixtures of ice 
and rock (b) and ice and iron (c). Curve (d) corresponds to a three component 
mixtures of ice, rock, and troilite, while curve (e) contains these three as 
well as organics and iron. Note that the the linear and  Garnett 
correlate very poorly for mixtures that contain even small mixing ratios of
elements with very high 
refractive indices. This message is also conveyed by figure \ref{fig:mix_garn} and the associated discussion. }
\label{fig:MG_lin}
\end{figure*}

Figure (\ref{fig:MG_lin}) shows a direct comparison of both 
the linear approximation and the generalized  Garnett function, using
the real and imaginary refractive indices of several combinations of likely 
materials at $\lambda=100 \,\mu$m as a function of the density ratio 
$\rho/\overline{\rho}$. Curve (a) corresponds 
to a 
single component of ice. Curves (b) and (c) are  
two-component mixtures of ice/rock and ice/iron. The fraction of iron 
is considerably less than the ice (see Table). Curve (d) 
is a three component mixture with the addition of troilite to ice and rock. 
Troilite also has a fairly large refractive index at this wavelength, which 
affects the linear approximation significantly. Curve (e) is a five component 
mixture with the contents of (d) augmented by iron and organics.  The  
Garnett dielectric constant 
is unaffected by the minute amount of iron, as has been found by others (Ossenkopf 1991), but the linear 
approximation diverges considerably because of iron's very high dielectric 
constant. This shows that metals, in general, 
are well beyond the realm of validity for the linear approximation.

The fractional mass and volume factors $\alpha_j$ and $f_j$ discussed above are simply related to 
the particle {\it number} density weighting factors $\beta_j$ of section 4 ({\it eg.} equation 16{\it ff}). Recall that $n(r)$ is the total number density of particles of all species, and if all the species have the same size distribution, the number density of particles of species $j$ is $n_j(r) = \beta_j n(r)$, 
where $\sum \beta_j = 1$. In the limit
where the scattering and absorption by each species needs to be separately 
treated (physically, for particles small enough to be monomineralic), 
$\rho_j(r) = \rho_j$ = constant. Then the total nebular mass density in species $j$ is 
\begin{equation}
\rho_{pj} = \int \frac{4}{3} \pi r^3 \rho_j \beta_j n(r) dr
    = \rho_j \beta_j \int \frac{4}{3} \pi r^3 n(r) dr
    = \rho_j \beta_j \xi,
\end{equation}
where $\xi$ is the nebula {\it volume  fraction} of all solids. Recall that $\rho_{p} = \sum \rho_{pj} = \alpha \rho_g$, where $\alpha$ is the total 
mass fraction solid material. Therefore,
\begin{equation}
    \beta_j = {\rho_{pj} \over \xi \rho_j }=  {\alpha_j \rho_g \over \xi \rho_j} = {\alpha_j \rho_g \overline{\rho} \over \rho_p \rho_j},
\end{equation}
where the last equality follows from $\rho_p \equiv \alpha \rho_g = \overline{\rho} \xi$, since for the heterogeneous particle case we assume the particle porosity $\phi=0$.  We previously defined the volume-averaged mass density of a compact particle as (\ref{eq:rhobar}):
\begin{equation}
\overline{\rho}=\frac{\alpha}{\sum_{j} (\alpha_{j}/\rho_{j})}
               = \frac{\rho_p}{\xi};
\end{equation}
then 
\begin{equation}\label{eq:betaj}
\beta_j = \frac{\alpha_j}{\rho_j} \frac{\rho_g}{\xi}
        = \frac{\alpha_j}{\rho_j} \frac{\rho_p}{\xi \alpha}
        = \frac{\alpha_j}{\rho_j} \frac{\overline{\rho}}{\alpha}
        = \frac{\alpha_j}{\alpha} \frac{\overline{\rho}}{\rho_j}
        = {\alpha_j \over \rho_j \sum_j (\alpha_j / \rho_j) }
        = \frac{f_j}{1-\phi},
\end{equation}
so it is simply verified that $\sum_j \beta_j = 1$. Moreover, equations (\ref{eq:rhobar}) and (\ref{eq:betaj}) can be combined to give $\alpha_j \overline{\rho} = \alpha \beta_j \rho_j$, and summing both sides over $j$ leads to the intuitively obvious alternative expression $\overline{\rho} = \sum_j \beta_j \rho_j$.
All of the above quantities are obtained from the presumed compositional makeup of 
the protoplanetary nebula.

\end{document}